\documentclass[aps,prx,twocolumn,superscriptaddress,nobibnotes]{revtex4-2}

\usepackage{graphicx}
\usepackage{dcolumn}
\usepackage{bm}
\usepackage{amsmath}
\usepackage{amssymb}
\usepackage{latexsym}
\usepackage{epsfig}
\usepackage{amsbsy}
\usepackage{array}
\usepackage{amssymb}
\usepackage{setspace}

    \usepackage[justification=centerlast]{caption}
    \usepackage{subfig}
    \usepackage{booktabs}

\begin{document}

\preprint{Physical Review Research}

\title{Multiple-scale integro-differential perturbation method for generic non-Markovian environments}

\author{Xiangyi~Meng}%
\email{xm@bu.edu}
\affiliation{Center for Polymer Studies and Department of Physics$\text{,}$ Boston University, Boston, Massachusetts 02215, USA}%
\author{Yang~Li}%
\affiliation{State Key Laboratory of Advanced Optical Communication$\text{,}$ Systems and Networks$\text{,}$ Department of Electronics$\text{,}$ and Center for Quantum Information Technology, Peking University, Beijing 100871, China}
\author{Jian-Wei~Zhang}%
\affiliation{School of Physics, Peking University, Beijing 100871, China}
\author{Hong~Guo}%
\affiliation{State Key Laboratory of Advanced Optical Communication$\text{,}$ Systems and Networks$\text{,}$ Department of Electronics$\text{,}$ and Center for Quantum Information Technology, Peking University, Beijing 100871, China}
\author{H.~Eugene~Stanley}
\email{hes@bu.edu}
\affiliation{Center for Polymer Studies and Department of Physics$\text{,}$ Boston University, Boston, Massachusetts 02215, USA}%

\date{\today}


\begin{abstract}

Non-Markovianity may significantly speed up quantum dynamics when the system interacts strongly with an infinite large reservoir, of which the coupling spectrum should be fine-tuned. The potential benefits are evident in many dynamics schemes, especially the continuous-time quantum walk. Difficulty exists, however, in producing closed-form solutions with controllable accuracy against the complexity of memory kernels. Here, we introduce a new multiple-scale perturbation method that works on integro-differential equations for general study of memory effects in dynamical systems. We propose an open-system model in which a continuous-time quantum walk is enclosed in a non-Markovian reservoir, that naturally corresponds to an error correction algorithm scheme. By applying the multiple-scale method we show how emergence of different time scales is related to transition of system dynamics into the non-Markovian regime. We find that up to two long-term modes and two short-term modes exist in regular networks, limited by their intrinsic symmetries. In addition to the effective approximation by our perturbation method on general forms of reservoirs, the speed-up of quantum walks assisted by non-Markovianity is also confirmed, revealing the advantage of reservoir engineering in designing time-sensitive quantum algorithms.
\end{abstract}

\maketitle

\hyphenpenalty=3000 
\tolerance=8000    
\hyphenation{Eq Fig ODP GME TCL}

\section{Introduction}

Many mathematical techniques have been developed to identify intrinsic time scales in dynamical systems. Multiple-scale perturbation~\cite{Holmes}, for example, is a highly-developed approximation method that solves complex dynamics in a perturbative way, by introducing trial variables as different time scales which are often of physical importance themselves. It is used in many fields, especially in such quantum optics topics~\cite{q-opt} as spontaneous radiation processes~\cite{multiple-scale-q-optics_llc73,multiple-scale-q-optics_ll73,multiple-scale_bgrvp19} and nonlinear solitons~\cite{multiple-scale-nonlinear-optics_lp84, multiple-scale-nonlinear-optics_b00}. A summary of its applications in quantum optical problems can be found in Ref.~\cite{multiple-scale-q-optics_j03}. Also originated from quantum optics is the study of open quantum systems~\cite{q-open-syst}, in which the dynamics of a quantum system is enriched when the system is open to an infinitely large reservoir from which microscopic interaction occurs. In particular, when the interaction spectrum is fine-tuned the open system may go through a quantum phase transition~\cite{q-phase-transit-nMarkov_nd15} and exhibit such non-Markovian features as bidirectional exchange of energy and coherence between the system and the reservoir, accompanied by singularity in time-dependent system variables~\cite{characterize-nMarkov_rhp14}. Assisted by non-Markovianity, unexpected atypical dissipation and decoherence behaviors~\cite{q-dissipative-syst} can arise in open systems, e.g., the sudden death of entanglement~\cite{entanglement-sudden-death_ye04,entanglement-sudden-death_talo10}. It
was recently found that non-Markovianity changes quantum speed limits~\cite{QSL_ceph13,QSL_dl13} and can significantly speed up quantum dynamics~\cite{QSL_mwg15}. The benefits of properly utilizing and
engineering non-Markovian reservoirs~\cite{reservoir-engineer_llhlglbp11,reservoir-engineer_klkcllmnh15} for designing time-sensitive quantum algorithms and protocols are obvious.

In this paper, we examine one specific quantum dynamics scheme, the \textit{continuous-time quantum walk}, of which the concept was first introduced in Ref.~\cite{q-walk_fg98}. As the name suggests, a walker moves continuously in time, navigating among different sites. Unlike a classical random walker, the propagation
of the quantum walker is \textit{coherent}, i.e., besides the randomness inherited from quantum-mechanical probability amplitudes the coherence between different sites also governs the system dynamics~\cite{q-walk-netw_mb11}. The study of continuous-time quantum walks is mathematically based on network theory~\cite{netw-introd,netw-of-netw_gbsh12,netw-spontaneous-recovery_mpbkhs14} and is closely related to other fields in quantum information theory, e.g., universal quantum computation~\cite{q-comput_c09}, quantum algorithms~\cite{q-walk-complete-graph_fg98,q-walk-star-graph_crpm18,q-walk-bipartite-graph_rw19}, and perfect state transfer~\cite{perfect-state-transfer_k10}. A continuous-time quantum walk can be experimentally implemented, usually by such quantum optical systems as waveguides~\cite{q-walk-waveguide_plpsms08} or Rydberg atoms~\cite{q-walk-rydberg_mbagrlw07}.

One of the breakthroughs in quantum information theory is the finding of a quadratical speed-up ($\sim O(\sqrt{N})$) in a quantum search algorithm, namely, Grover's algorithm~\cite{grover_g97}. Grover's algorithm has proven to be equivalent to a continuous-time quantum walk on a complete network~\cite{q-walk-complete-graph_fg98}. In addition, a speed-up by continuous-time quantum walks on a star network has also been found~\cite{q-walk-star-graph_crpm18}, and this has encouraged further study
of continuous-time quantum walks on regular networks. Here, a network is said to be regular if it has a repeating pattern in its network topology. This should not be confused with $k$-regular graphs which are defined differently. To expand the theoretical structure, research on open-system quantum walks has also been conducted~\cite{q-walk-open-syst_apss12,q-walk-open-syst_p14} in order to deal with noisy environments~\cite{q-walk-nMarkov_bbbp16,q-walk-dyn-percolate_brp19}, but exact solutions or model-based descriptions are rare, especially in the non-Markovian regime where the system dynamics is inevitably governed by strong memory effects. On the other hand, we expect that many interesting features granted by non-Markovianity should also exist in the quantum walk scheme, among which the most useful is the steep decrease of quantum speed limit~\cite{QSL_mwg15}.

To construct a practical methodology and verify the non-Markovianity-assisted speed-up, we propose a generalized version of the multiple-scale perturbation method that now can work on integro-differential equations. Strong memory effects can be studied by perturbatively expanding the memory kernel. The advantage of using a perturbation method is being able to derive closed-form approximate solutions where accuracy and complexity are controllable. We further apply the perturbation method to a continuous-time quantum walk enclosed in a non-Markovian reservoir. Such a model naturally follows an error correction algorithm scheme, with the reservoir a collection of independent ``error'' sites. We find that two time scales of different physical importance emerge when the studied system moves into the non-Markovian regime. With the accuracy of our perturbation method being guaranteed, we investigate how quantum-walk dynamics is affected by the coupling strength between system and reservoir, as well as by the intrinsic network topology. The expected speed-up is confirmed by looking into the four different eigenfrequencies hidden in the non-Markovian dynamics, inherited from symmetries of regular networks. 

It is unclear yet whether this speed-up can be utilized as a quantum resource, given that non-Markovianity can be detrimental for certain delicate quantum tasks~\cite{q-walk-nMarkov-detrimental_rcpm18}. We expect that non-Markovianity should be the most useful for tasks that are speed-focused, e.g., quantum simulation~\cite{q-simul-open-syst_mh09}. Our method is suitable to do a more general study on this issue.

The rest of this paper is organized as follows. Section~\ref{section_math} introduces the multiple-scale perturbation method and its generalization to integro-differential equations, which we expect to have broader applications than quantum walks. Section~\ref{section_ctqw} introduces the concept of continuous-time quantum walks and the quantum algorithm picture behind the interaction with non-Markovian reservoirs. Section~\ref{section_n-markov} presents a test of the performance of our multiple-scale approximation method for general reservoirs and examines the relation between time scales and non-Markovianity. Section~\ref{section_regular-netw} describes continuous-time quantum walks on some regular networks, including complete networks, star networks, rings, and square lattices, and reveals the existence of up to four system modes in terms of the two time scales and how they connect to network topology.

Remaining questions in this crossover study of non-Markovian memory effect and continuous-time quantum walks include the possibility that there are more than two independently important time scales in the non-Markovian model. We hope to understand these time scales better in a systematic manner in the future. We could also apply our method to quantum walks on complex networks~\cite{q-walk-complex-netw_fjbkm13} where the statistics of disorder could take unpredictable new forms.

\section{Multiple-scale integro-differential perturbation method}
\label{section_math}
Often in dynamical systems there is no exact solution $y(x)$ to the system dynamics. It is difficult to acquire accurate and reliable approximations. The regular perturbation method~\cite{Holmes} is a powerful approach in finding an approximation of the unknown dynamics. By introducing a small dimensionless perturbation parameter $\alpha$, any $y(x)$ can be expressed in terms of a power series in $\alpha$ and can be well approximated in a closed form by finite leading terms. In practice, however, given an intricate system, the complex behavior of $y(x)$ often invalidates the perturbation approximation. In the power series, each term follows the same approximate over-simplified dynamics, causing the complexity of $y(x)$ to be limited. Thus the complex behavior is unavoidably lost when the infinite series is broken down.

The \emph{multiple-scale perturbation method}~\cite{Holmes} has then been introduced to overcome this difficulty. The basic idea is to add more degrees of freedom in terms of new independent variables into the system. Different characteristics are captured by different variables, an attempt to retrieve the complexity of $y(x)$ in each single perturbation term. The procedure of the multiple-scale perturbation method goes as follows: first, two (or more) different and artificial scales are chosen as functions of $x$ and $\alpha$; the scales are considered as new independent variables, allowing the ordinary differential equations to be converted into partial differential equations and expanded into perturbation series. The additional degree of freedom from the extra variables is eventually constrained by requiring that higher perturbation terms diverge no more quickly than lower perturbation terms~\cite{Holmes}.

The classical Duffing equation
\begin{equation*}
\frac{d^2}{dx^2}y+y+4\alpha y^3=0
\end{equation*}
is a neat example appearing in the study of anharmonic oscillators, which cannot be solved by a regular perturbation method due to its small nonlinear term~\cite{multiple-scale-q-optics_j03}. In fact, the secular term in the first-order perturbation correction $y^{\left(1\right)}\left(x\right)$ is proportional to $x \sin \left(x\right)$, which is unbounded and incorrect~\cite{multiple-scale-q-optics_j03}. Now, by choosing two independent scales, $u=x$ and $v=\alpha x$, one has, locally,
\begin{equation*}
\frac{d}{dx}=\frac{\partial }{\partial u}+\alpha\frac{\partial }{\partial v}
\end{equation*}
in the neighborhood of $x$. A term-by-term perturbation expansion yields
\begin{eqnarray*}
	\frac{\partial^2}{\partial u^2}y^{\left(0\right)}+y^{\left(0\right)}&=&0,\\
	\frac{\partial^2}{\partial u^2}y^{\left(1\right)}+y^{\left(1\right)}&=&-4 \left(y^{\left(0\right)}\right)^3-2\frac{\partial^2}{\partial u \partial v}y^{\left(0\right)},
\end{eqnarray*}
up to the first two orders. Thus one has $y^{\left(0\right)}(u,v)=h(v)\cos u+k(v)\sin u$. $h(v)$ and $k(v)$ are determined next by requiring $-4 \left(y^{\left(0\right)}\right)^3-2\frac{\partial^2}{\partial u \partial v}y^{\left(0\right)}=0$ so that $y^{\left(1\right)}$ does not diverge. With initial conditions $y(0)=1$ and $y'(0)=0$ the final result reads
\begin{equation*}
y^{\left(0\right)}=\cos \left[\left(1+3\alpha/2 \right)x\right],
\end{equation*}
which is not divergent and exhibits the first-order correction to the frequency~\cite{multiple-scale-q-optics_j03}.

Based on the same thought, we generalize the method to an integro-differential equation which is of a general form
\begin{equation}
\label{integro-differential}
\mathcal{F}(\frac{d}{dx},y(x))+\int_{0}^{x}{dx'G(x-x')y(x')}=0.
\end{equation}
Here, $\mathcal{F}(d/dx,y(x))$ represents an arbitrary differential term(s) and $G(x-x')$ is a convolution kernel. The difficulty arises when an integral term is added, that the convolution is not a local operation and thus the artificial scales cannot be considered independent. The integral term has to be dealt with indirectly. If the kernel is holomorphic near $x=x'$, then $G(x-x')$ can be expanded as
\begin{equation}
	\label{kernel-expansion}
	\alpha^p \left[G_0 +  G_1 \alpha^{q} \left( x-x' \right) +  G_2 \alpha^{2q} \left( x-x' \right)^2+\cdots\right],
\end{equation}
a series of $(x-x')$ living in its neighborhood without any singularity. $p$ and $q$ are understood as integers, w.l.o.g. The expansion further suggests that locality can be regained from Eq.~(\ref{integro-differential}) if the perturbation procedure is fine-tuned. Here, the trick is to bring Eq.~(\ref{integro-differential}) into higher and higher differential orders, meanwhile trying to cancel out integral terms or at least make them comparably smaller than differential terms. The iterative procedure of the multiple-scale perturbation method thus contains three steps for each perturbation order:
\begin{enumerate}
	\item Let $d/dx$ act on both sides of the integro-differential equation to solve. Recall that $(d/dx)^{n+1}\int{d{x'}(x-x')^{n}y(x')}=n!y(x)$, a guarantee of being locally holomorphic near $x$.	
	\item Introduce artificial scales and replace $d/dx$ by partial differential operators in terms of the new variables.
	\item Extract the lowest order terms. They are the corresponding perturbation correction to be calculated. The rest terms are saved for next iteration.
\end{enumerate}
During the iteration, more and more integral terms are differentiated. With a successful choice of scales and $\alpha$, all perturbation corrections should only consist of local terms, of which the solutions are easily carried out by the usual multiple-scale approach~\cite{Holmes}.

Note that perturbation methods enable us to understand the functional importance of different dynamic terms and parameters. The solutions generated by perturbation methods are also in a closed form that is more suitable for practical purposes, in part because the precision of numerical calculations in closed-form functions are more controllable than in complex functions.  The study of integro-differential equations has traditionally involved the use of integral transforms, but few solutions have closed-form expressions. We will see that the regular perturbation method also cannot be used if we are to expand the solutions perturbatively in the transformed domain. The different behaviors of intrinsic scales in the system simply cannot be captured, the perturbation corrections remaining unbounded or erratic. Hence the use of the multiple-scale perturbation method is essential to meet the exact needs.

\section{Continuous-time quantum walks}
\label{section_ctqw}
The simplest definition of a continuous-time quantum walk (CTQW) involves a connected network $\mathcal{G}(V,E)$ which comprises a set $V$ of nodes and a set $E$ of links. The weighted adjacency matrix $\mathbf{A}$ of $\mathcal{G}$ is an Hermitian matrix of dimension $N=\left|V\right|$. The matrix elements of $\mathbf{A}$ satisfy $A_{ij}=A^*_{ij}$. $A_{ij}\equiv0$ when $(i,j)\notin E$. The walker is a $N$-dimensional complex vector $\mathbf{c}(t)$ which follows the dynamics
\begin{equation}
\label{ctqw}
\dot{\mathbf{c}}+i \mathbf{A} \mathbf{c}=0.
\end{equation}

Sometimes, instead of an adjacency matrix, a Laplacian matrix is preferred, yet for both the dynamics is equivalent on regular networks~\cite{q-walk-laplacian-vs-adjacency_wtn16}. The first systematic study of Eq.~(\ref{ctqw}) from a statistical physics perspective dates back to the Anderson localization model~\cite{Anderson-local_a58}, where the diffusion of a single electron that scatters in-between $N$ sites, if described well by short-range interaction and tight-binding approximation, can be reduced to simpler dynamics in the form of Eq.~(\ref{ctqw}). The diffusion behavior in terms of the electron state $\left|\psi(t)\right>=\sum_{i}c_i(t)\left|i\right>$ is determined by the amount of ``impurity'' in the Hamiltonian $H$. Between any two sites $i$ and $j$ the interaction $\left<i\right|H\left|j\right>$ is simply $A_{ij}$. 

Another perspective---which is more general---is to consider a spin system with Hamiltonian $H_{\mathcal{G}}=\sum_{i,j}(A_{ij}\sigma^i_{+}\sigma^j_{-}+A_{ji}\sigma^j_{+}\sigma^i_{-})$ where the occupation number $\sum_{i}\sigma^i_{+}\sigma^i_{-}$ in the Fock space is manifestly conserved, and view the continuous-time quantum walk as a spin diffusion in the one-exciton Hilbert subspace $\mathcal{H}_{\text{sub}}=\mathbb{C}^N$. The operators ${{\sigma }_{\pm }}$ are Pauli operators. We note that statistical studies of such a spin system beyond one exciton have undergone difficulties and eventually led to the theory of many-body localization~\cite{many-body-local_baa06}, which is however not our focus here.

\subsection{Quantum walk in a reservoir}

\begin{figure}[h]
	\includegraphics[width=8.6cm]{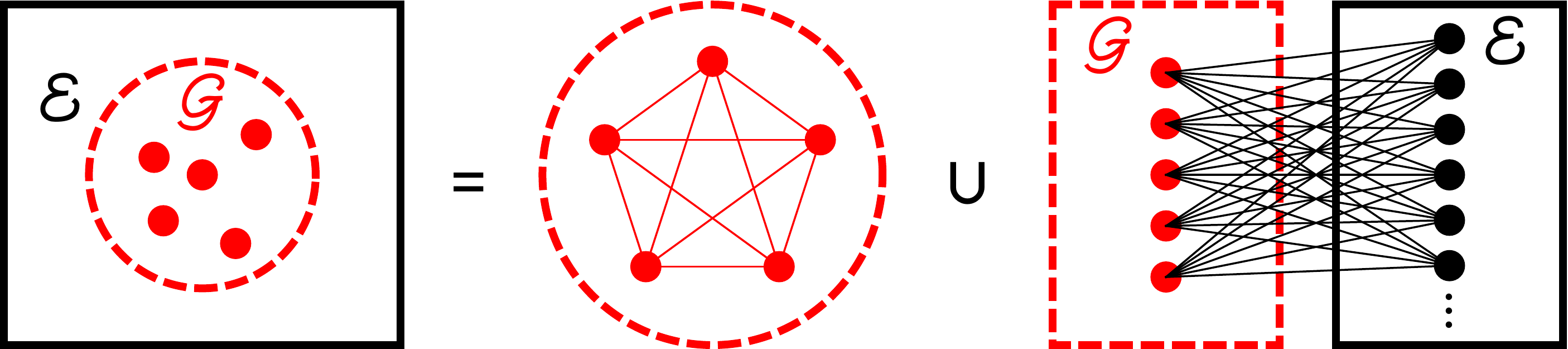}
	\caption{\label{fig_diagram}(Color online) A continuous-time quantum walk on an arbitrary network $\mathcal{G}(V,E)$, embraced by a reservoir $\mathcal{E}$. The quantum dynamics on $\mathcal{G}$ is an example of open quantum system dynamics~\cite{q-open-syst}. The whole system can also be viewed as the union of two networks: an independent network $\mathcal{G}$ and a complete bipartite network $(V,V_{\mathcal{E}},\forall)$ where $V_{\mathcal{E}}$ is infinitely large.\hfill \hfill}
\end{figure}

The solution of Eq.~(\ref{ctqw}) is relatively simple and has been studied well even in the $N\to\infty$ limit~\cite{Anderson-3d_ba96}. However, by introducing a large reservoir that is described by a free Hamiltonian with infinite modes, $H_{\mathcal{E}}=\sum_{k=1}^{\infty}\varepsilon_k a^\dagger_k a_k$, the system dynamics will be dramatically changed if further undergoing an interaction term, 
\begin{equation*}
H_{\text{int}}=\sum\nolimits_{i,k}\left(w_i g_k\right) \sigma^i_{+}a_k+ \text{h.c.},
\end{equation*}
where $w_i g_k$ is the coupling between spin $i$ and mode $k$, and ${a}_{k}$ and $a_{k}^{\dagger }$ are annihilation and creation operators of either bosons or fermions. The subspace $\mathcal{H}_{\text{sub}}=\mathbb{C}^N\oplus\mathbb{C}^{\infty}$ is still independent though, conserved by $\sum_{i}\sigma^i_{+}\sigma^i_{-}+\sum_{k} a^\dagger_k a_k=1$, where it does not matter whether ${a}_{k}$ and $a_{k}^{\dagger }$ follow commutation or anti-commutation relations. The quantum walker state can be written as  $\left|\psi(t)\right>=\sum_{i}c_i(t)\left|1_i\right>_{\mathcal{G}}\left|0\right>_{\mathcal{E}}+\left|0\right>_{\mathcal{G}}\left|\chi(t)\right>_{\mathcal{E}}$ where $\chi(t)$ is some unknown distribution in $\mathbb{C}^{\infty}$, the Fock subspace of $\mathcal{E}$. The dynamics of $\left|\psi(t)\right>$ is governed by $H_{\mathcal{G}}+H_{\mathcal{E}}+H_{\text{int}}$ together, and $c_i(t)$ itself is described by open system dynamics~\cite{q-open-syst}. After some calculations we derive, in the matrix formulation, 
\begin{equation}
\label{integro-differential_ctqw}
\dot{\mathbf{c}}(t)+i \mathbf{A}\mathbf{c}(t)=-\int_{0}^{t}{dt'\mathbf{G}(t-t')\mathbf{c}(t')},
\end{equation}
or simply $\dot{\mathbf{c}}+i \mathbf{A}\mathbf{c}=-\mathbf{G}\ast\mathbf{c}$ where $\ast$ is the convolution operation. 
We have
\begin{equation*}
\mathbf{c}=\left(\begin{matrix}
c_1\\
c_2\\
\vdots\\
c_N
\end{matrix}\right),\qquad\qquad
\mathbf{w}=\left(\begin{matrix}
w_1\\
w_2\\
\vdots\\
w_N
\end{matrix}\right),
\end{equation*}
and $\mathbf{G}(t)=\mathbf{w}\left(\mathbf{w}^*\right)^{\intercal} G(t)$, the memory kernel $G$ being
\begin{equation}
\label{G}
G(t)=\sum\nolimits_{k}{{{g}_{k}}g_{k}^{*}{{e}^{-i \varepsilon_k t}}}\simeq \int{d\omega }{{J}}(\omega){{e}^{-i\omega t}}.
\end{equation}
The Fourier transform of $G(t)$ is called the \textit{spectral density} $J(\omega)$ of the reservoir~\cite{q-open-syst}. When $J(\omega)$ is close to a constant, the spectrum is white-noise like, yielding a memory-less kernel $G(t)\sim\delta(t)$. The open system is Markovian and is only subjected to exponential decay. However, predicted by the open quantum system theory, a ``colored'' spectrum will cause non-Markovian effect, and in the extreme case $J(\omega)\sim\delta(\omega)$ the system undertakes to-and-fro oscillation without sign of dissipation. In general, $\mathbf{G}(t)$ does not have to be a single form. If each element $G_{ij}(t)$ has independent spectrum density, then the system behavior will be even more complex.

Thus different forms of $J(\omega)$ suggest completely different behaviors~\cite{structured-optical-reservoirs_lnnb00}. We therefore assume that there is a universal dimensionless factor $\alpha$ which compares the bandwidth of $J(\omega)$ to its peak. For multimodal distributions $\alpha$ is interpreted as the average for different modes. In the \emph{weak-coupling regime}, $\alpha\gg1$, the Markovian open system is well described by the \textit{Born-Markov approximation}~\cite{q-dissipative-syst}, a regular perturbation approach with the perturbation parameter being $\alpha^{-1}$. In the \emph{strong-coupling regime}, $\alpha\ll1$, the non-Markovian system behaves more interestingly, but the regular perturbation method no longer works~\cite{q-open-syst}.

Note that this model can be easily realized in cavity QED~\cite{cavity-QED_hk89}, where $H_{\mathcal{G}}$ describes a system of two-level dipoles, $H_{\mathcal{E}}$ describes a quantized radiation bath with $a_k$ and $a^{\dagger}_k$ being electromagnetic fields, and $w_i g_k$ represents the strength of coupling between dipole $i$ and cavity mode $k$. The quality factor of cavity $Q$ is important~\cite{q-opt} and can be related to the perturbation parameter by $\alpha\sim Q^{-1/2}$. It is of practical interest as well, that all results in this paper can actually be tested through cavity-QED experiments.

Back to network theory, as shown in Fig.~\ref{fig_diagram}, the open system and reservoir can be considered together as the union of $\mathcal{G}(V,E)$ and a complete bipartite network $(V,V_{\mathcal{E}},\forall)$. Here the (mathematical) union $\cup$ of two networks $\mathcal{G}_1(V_1,E_1)$ and $\mathcal{G}_2(V_2,E_2)$ yields a new network $\mathcal{G}_{1\cup2}(V_1\cup V_2,E_1\cup E_2)$. $V_{\mathcal{E}}$ is the set of the infinite nodes in $\mathcal{E}$. The network $(V,V_{\mathcal{E}},\forall)$ here is bipartite, i.e., it has two disjoint and independent set of nodes, $V$ and $V_{\mathcal{E}}$, and has links of which each only connects one node from $V$ to one node from $V_{\mathcal{E}}$. The symbol $\forall$ denotes that the bipartite network actually contains every link from $V$ to $V_{\mathcal{E}}$ and thus is complete. 

It is a conceptually useful point of view, to regard all nodes that ${\mathcal{E}}$ consists of as ``error'' sites. By walking into an ``error'' site, the walker realizes its mistake and attempts to go back to the possible ``target'' sites in $\mathcal{G}$, but never to enter another ``error'' site. Therefore, for any quantum search algorithm in terms of CTQW, the dynamics here naturally follows an error correction scheme that functions on a solution space $\mathcal{G}$ plus an ``error'' space $\mathcal{E}$. Within this scheme, it is clear that the error correction is carried out by non-Markovian feedback controls, the efficiency of which is dependent not only on the ``error'' rate $\mathbf{w}$ but also on the spectrum, $J(\omega)$. This leads us to better understanding of error correction algorithm design.

\subsection{Perturbation expansion}
The last step is to apply the multiple-scale integro-differential perturbation method to Eq.~(\ref{integro-differential_ctqw}). We introduce three new dimensionless quantities, $\tilde{t}=\gamma t$, $\mathbf{\tilde{A}}=\gamma^{-1} \alpha^{-1} \mathbf{A}$, and $\mathbf{\tilde{G}}=\gamma^{-2} \alpha^{-2} \mathbf{G}$. Here $\gamma\sim\sum\nolimits_{i,k}w_i g_k$ has a dimension of inverse time. We choose two independent time scales, a primary scale $T$ and an auxiliary scale $\tau$, written as
\begin{eqnarray}
\label{tauT}
T&=&\left( {{A}_{0}}\alpha\left.\right. +{{A}_{1}}{{\alpha }^{3}}+\cdots +{{A}_{n}}{{\alpha }^{2n+1}}+\cdots  \right)t,\nonumber\\
\tau &=&\left( {{B}_{0}}\alpha^2 +{{B}_{1}}{{\alpha }^{4}}+\cdots +{{B}_{n}}{{\alpha }^{2n+2}}+\cdots  \right)t,
\end{eqnarray}
and also, $\tilde{T}=\gamma T$ and $\tilde{\tau}=\gamma\tau$. The coefficients $\{A_n\}$ and $\{B_n\}$ are to be determined. We assume that $\mathbf{\tilde{G}}(\tilde{t})$ is holomorphic near $\tilde{t}$, and $\mathbf{\tilde{G}}(\tilde{t})=\mathbf{w}\left(\mathbf{w}^*\right)^{\intercal}\sum_{n}G_n\alpha^{2n}\tilde{t}^n$. Finally, we have
\begin{eqnarray}
\label{integro-differential_ctqw_expand}
&&\sum_{n,m}\left[\alpha^{2n}\left(A_n \alpha\frac{\partial}{\partial \tilde{T}}+B_n \alpha^{2}\frac{\partial}{\partial \tilde{\tau}}\right)\mathbf{I}+i\alpha\mathbf{\tilde{A}}\right] {{\alpha }^{m}\mathbf{c}^{\left(m\right)}(\tilde{T},\tilde{\tau} )}\nonumber\\
&&=-\sum_{n,m}\int_{0}^{\tilde{t}}{d{\tilde{t}'}{G_{n}{\alpha }^{2+2 n+m}\left(\tilde{t}-\tilde{t}'\right)^{n}}\mathbf{w}\left(\mathbf{w}^*\right)^{\intercal}{\mathbf{c}^{\left(m\right)}(\tilde{t}')}}\nonumber\\
\end{eqnarray}
derived from Eq.~(\ref{integro-differential_ctqw}), with $\mathbf{c}(\tilde{t})=\sum_m\alpha^m\mathbf{c}^{\left(m\right)}(\tilde{t})$ and $\mathbf{I}$ the identity matrix. Note that $\tilde{T}$, $\tilde{\tau}$, and $\tilde{t}$ appear simultaneously in Eq.~(\ref{integro-differential_ctqw_expand}), a result of mixing local partial derivatives and nonlocal integrals together.

Following the perturbation procedure, we let $d/d\tilde{t}$ act on both sides of Eq.~(\ref{integro-differential_ctqw_expand}) and extract the terms of the lowest perturbation order,
\begin{eqnarray*}
	\alpha^2 A_0^2 \frac{\partial^2}{\partial\tilde{T}^2} \mathbf{c}^{\left(0\right)}+i \alpha^2 A_0 \mathbf{\tilde{A}}\frac{\partial}{\partial\tilde{T}} \mathbf{c}^{\left(0\right)} =-\alpha^2 G_0 \mathbf{w}\left(\mathbf{w}^*\right)^{\intercal}\mathbf{c}^{\left(0\right)},\nonumber\\
\end{eqnarray*}
which yields,
\begin{equation}
\label{integro-differential_ctqw_expand_0}
	\left(\begin{matrix}
	0&\mathbf{I}\\
	-G_0 \mathbf{w}\left(\mathbf{w}^*\right)^{\intercal}&-i\mathbf{\tilde{A}}
	\end{matrix}\right)
	\left(\begin{matrix}
	\mathbf{c}^{\left(0\right)}\\
	\partial\mathbf{c}^{\left(0\right)}
	\end{matrix}\right)
	=
	\frac{\partial}{\partial\tilde{T}}
	\left(\begin{matrix}
	\mathbf{c}^{\left(0\right)}\\
	\partial\mathbf{c}^{\left(0\right)}
	\end{matrix}\right).
\end{equation}
The solution of Eq.~(\ref{integro-differential_ctqw_expand_0}) is
\begin{equation*}
\mathbf{c}^{\left(0\right)}(\tilde{T},\tilde{\tau})=\sum\nolimits_{s}\mathbf{f}_{s}^{\left(0\right)}(\tilde{\tau})e^{-i \tilde{\omega}_s^{\left(0\right)} \tilde{T}}
\end{equation*}
where $\tilde{\omega}_s^{\left(0\right)}$ are the eigenvalues of the matrix on the left-hand side of Eq.~(\ref{integro-differential_ctqw_expand_0}). For the sake of consistency we introduce $\omega_s^{\left(0\right)}=\tilde{\omega}_s^{(0)}\gamma$. $\mathbf{f}_{s}^{(0)}(\tilde{\tau})$ are to be determined from the equation of the next perturbation order,
\begin{eqnarray*}
&&\alpha^4 A_0^3 \frac{\partial^3}{\partial\tilde{T}^3} \mathbf{c}^{\left(1\right)}+
2\alpha^4 A_0^2 B_0 \frac{\partial^3}{\partial\tilde{T}^2\partial\tilde{\tau}} \mathbf{c}^{\left(0\right)}\nonumber\\
&&+i \alpha^4 A_0^2 \mathbf{\tilde{A}}\frac{\partial^2}{\partial\tilde{T}^2} \mathbf{c}^{\left(1\right)}+i \alpha^4 A_0 B_0 \mathbf{\tilde{A}}\frac{\partial^2}{\partial\tilde{T}\partial\tilde{\tau}} \mathbf{c}^{\left(0\right)}\nonumber\\
&=&-\alpha^4 A_0 G_0 \mathbf{w}\left(\mathbf{w}^*\right)^{\intercal}\frac{\partial}{\partial\tilde{T}}\mathbf{c}^{\left(1\right)}-\alpha^4 G_1 \mathbf{w}\left(\mathbf{w}^*\right)^{\intercal}\mathbf{c}^{\left(0\right)},
\end{eqnarray*}
derived by letting $d/d\tilde{t}$ act again on the rest terms and then extracting the new lowest-order ones. Note that $\mathbf{c}^{\left(1\right)}$ terms should not dominate $\mathbf{c}^{\left(0\right)}$ terms, so all $\mathbf{c}^{\left(0\right)}$ must cancel out with each other, which further implies
\begin{equation}
\label{integro-differential_ctqw_expand_1}
G_1\left[2\left(-i \tilde{\omega}_s^{\left(0\right)}\right)^2\mathbf{I}+ \tilde{\omega}_s^{\left(0\right)}\mathbf{\tilde{A}}\right]^{-1}\mathbf{w}\left(\mathbf{w}^*\right)^{\intercal}\mathbf{f}_{s}^{\left(0\right)}=\frac{\partial}{\partial\tilde{\tau}}\mathbf{f}_{s}^{\left(0\right)}
\end{equation}
where $\mathbf{f}_{s}^{\left(0\right)}$ are of exponential form, $\mathbf{f}_{s}^{\left(0\right)}\propto\exp(-\tilde{\Gamma}_s^{(0)} \tilde{\tau})$, which are to be fixed by initial conditions. $\Gamma_s^{(0)}=\tilde{\Gamma}_s^{(0)}\gamma$ are the decay rates.

Higher-order perturbation corrections are subjected to the same procedure, where $\{A_n\}$ and $\{B_n\}$ are also going to be fixed, an example of which is given in the next section.

\section{non-Markovian reservoirs}
\label{section_n-markov}
In this section, we study $N=1$ systems and investigate different non-Markovian reservoirs in order to understand how non-Markovianity changes system behavior and invalidates the regular perturbation method.
\subsection{Example: Lorentzian reservoir}
A Lorentzian reservoir is one of the few types that can be solved exactly in closed form, which is the reason why it is used here as a preliminary example. The spectral density is
\begin{equation*}
\label{JwLorentz}
J(\omega )=\frac{1}{2\pi}\frac{{{\gamma }}{{\lambda }^{2}}}{ {{\left(\omega- \Delta  \right)}^{2}}+{{\lambda }^{2}} },
\end{equation*}
where $\lambda$ is the spectral width, ${\gamma}$ is the coupling strength, and $\Delta$ is the off-resonance frequency. The corresponding memory kernel is
\begin{equation}
\label{Gexp}
G(t-t')=\frac{1}{2} \gamma \lambda {e}^{-\left(\lambda+i\Delta\right)\left(t-t'\right)}.
\end{equation}
It is clear that Eq.~(\ref{integro-differential_ctqw}) has a closed-form exact solution by Laplace transform (LT) because of the exponential form of Eq.~(\ref{Gexp}).
	
\begin{figure}[t]
	\includegraphics[width=8.6cm]{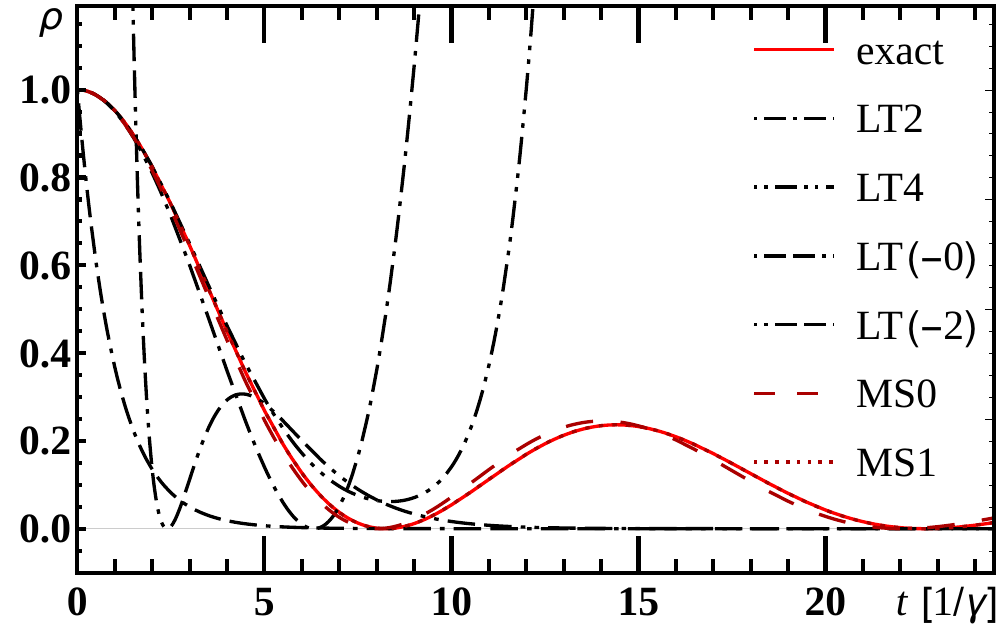}
	\caption{\label{fig_single}(Color online) State population of a single-node system in a Lorentzian reservoir. Here $\lambda =\gamma/10$, indicating that the system is in the strong-coupling regime. Approximate solutions by the regular perturbation method on the $s$ domain of Laplace transform (black) fail to match the exact solution (red), no matter to which positive or negative order the approximations being calculated. As comparison, closed-form approximate solutions by the multiple-scale perturbation method (brown) to the first two orders are already in good agreement with the exact solution.\hfill \hfill}
\end{figure}

\begin{figure*}[t]
	\begin{centering}
		
		\subfloat[ ]{\begin{centering}
				\label{fig_reservoirs_a}
				\includegraphics[width=5.7cm]{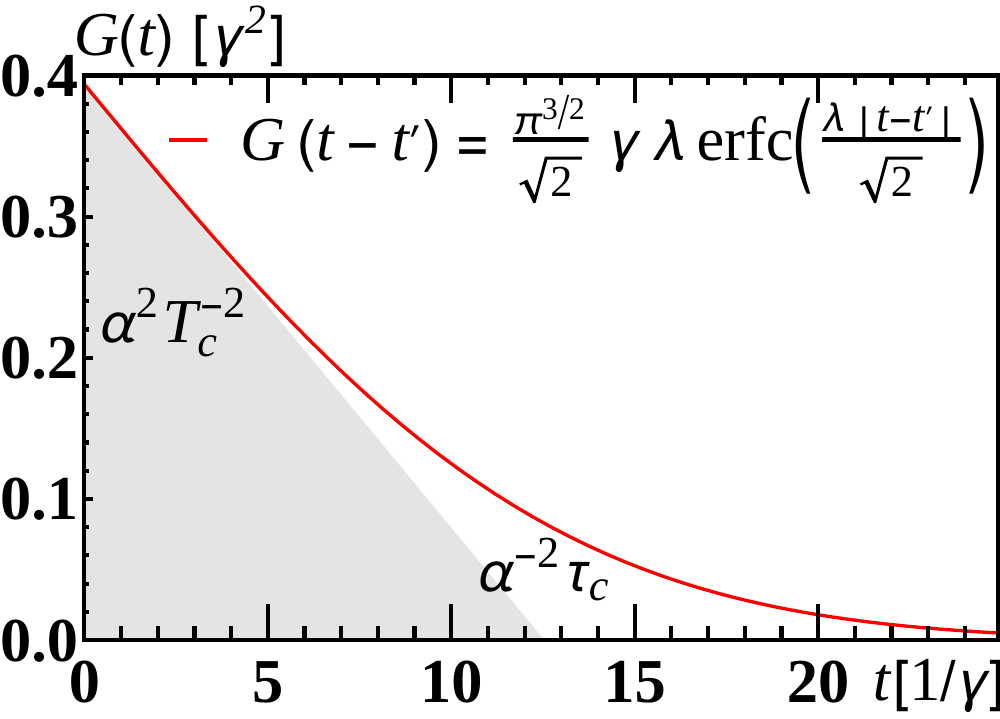}
				\par\end{centering}
		}\subfloat[ ]{\begin{centering}
			\label{fig_reservoirs_b}
			\includegraphics[width=5.7cm]{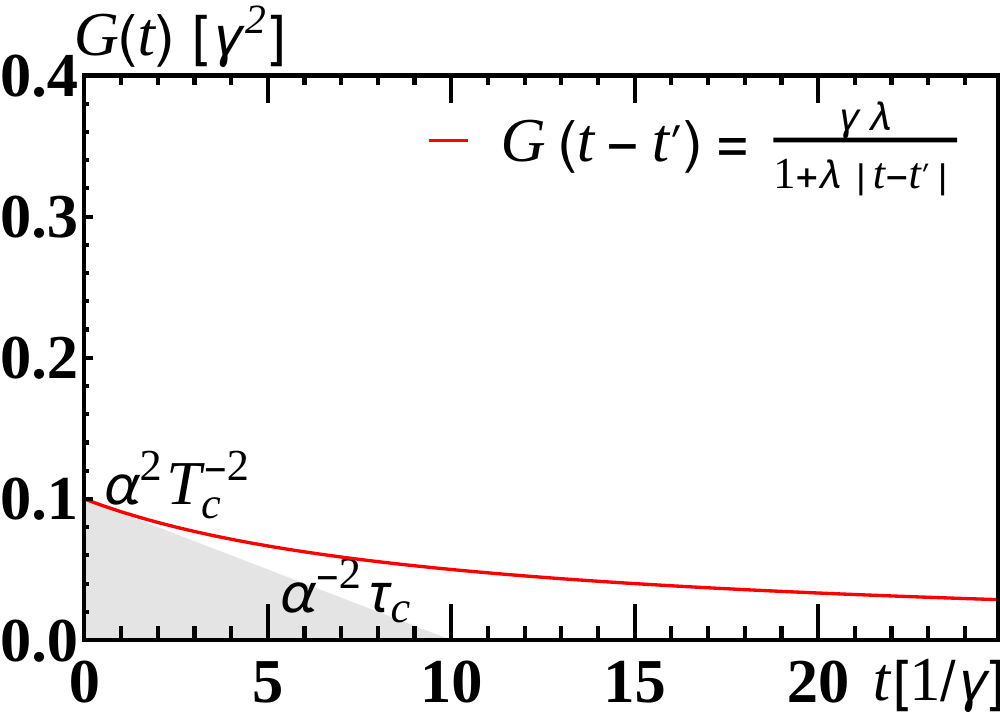}
			\par\end{centering}
	}\subfloat[ ]{\begin{centering}
		\label{fig_reservoirs_c}
		\includegraphics[width=5.7cm]{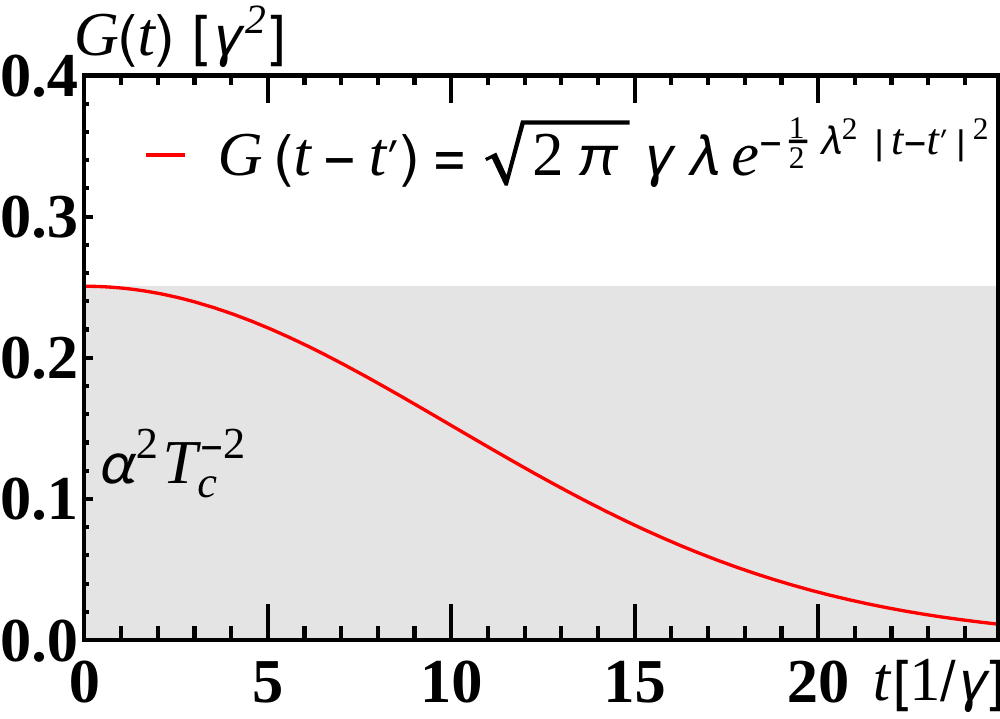}
		\par\end{centering}
}

\subfloat[ ]{\begin{centering}
		\label{fig_reservoirs_d}
		\includegraphics[width=5.7cm]{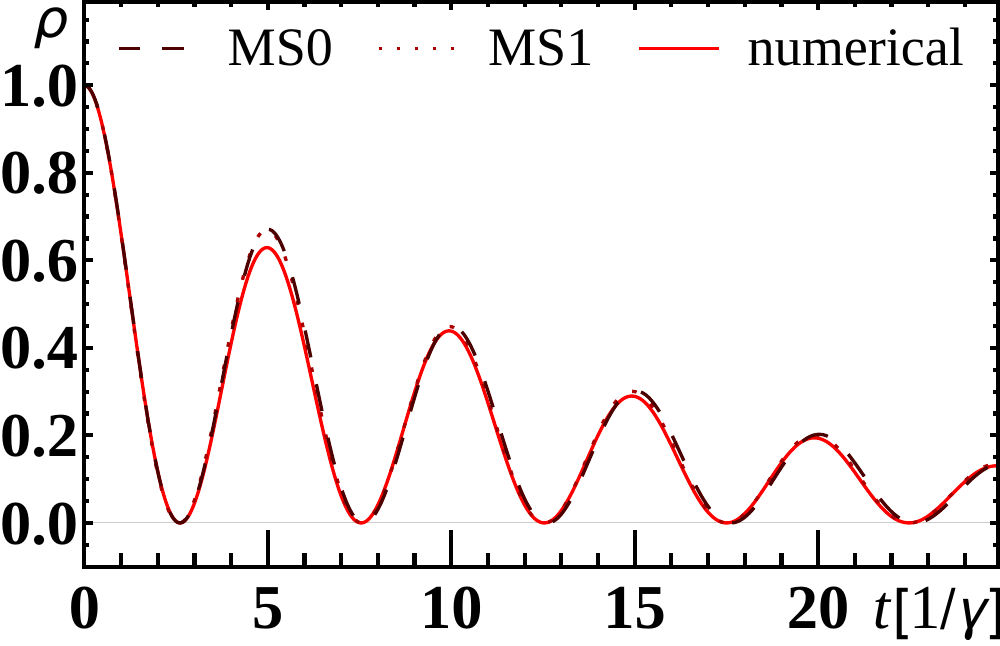}
		\par\end{centering}
}\subfloat[ ]{\begin{centering}
	\label{fig_reservoirs_e}
	\includegraphics[width=5.7cm]{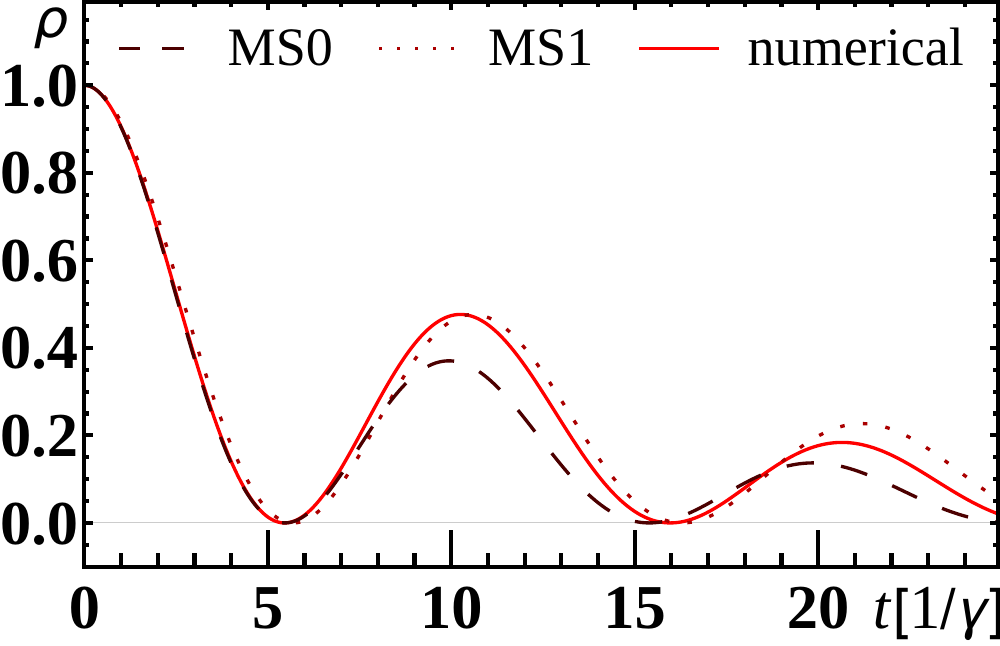}
	\par\end{centering}
}\subfloat[ ]{\begin{centering}
	\label{fig_reservoirs_f}
	\includegraphics[width=5.7cm]{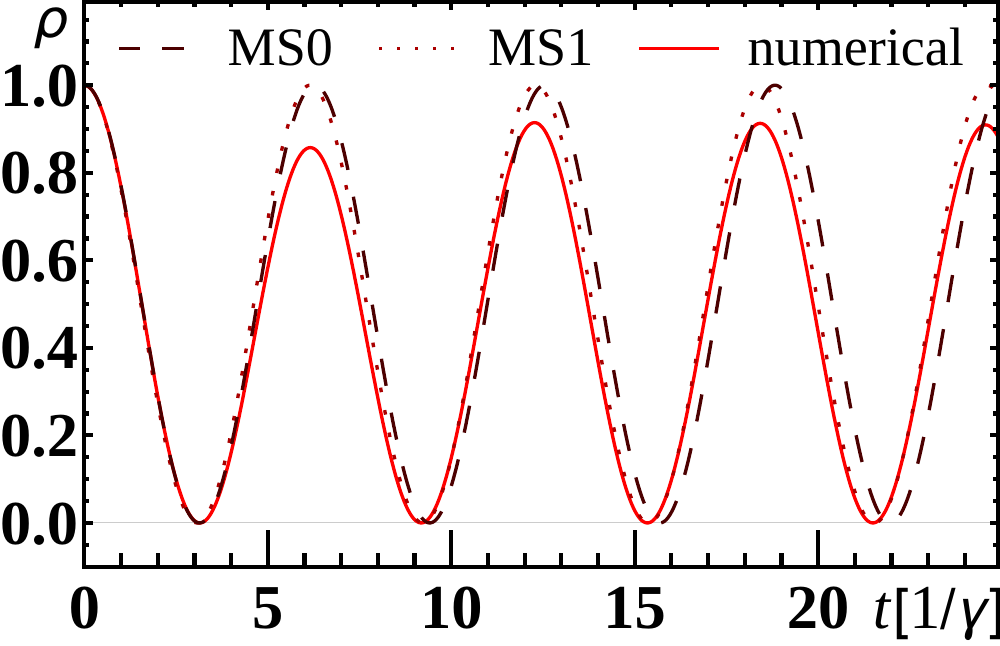}
	\par\end{centering}
}

\par\end{centering}
\caption{\label{fig_reservoirs}(Color online) Non-Markovian reservoir characterized by a memory kernel which has the form of a (a), Gauss error distribution, (b), inverse-law distribution, and (c), Gaussian distribution, from which the characteristic times $T_c$ and $\tau_c$ are locally determined (shaded area). (d)-(f): Closed-form approximate solutions of the state population of a single-node system in the corresponding reservoirs by the multiple-scale perturbation method (brown), compared to the numerical solution (red). $\lambda =\gamma/10$.\hfill \hfill}
\end{figure*}

\subsubsection{Comparison of perturbation methods}
We only consider the simplest configuration that $\Delta=0$ so the system is on resonance. The solution of Eq.~(\ref{integro-differential_ctqw}) is simply
\begin{equation}
\label{lorentz_exact}
c(t)={{e}^{-\frac{1}{2}\lambda t}}\left[ \cos \left( \frac{1}{2}D t \right)+\frac{\lambda}{D}\sin \left( \frac{1}{2}D t \right) \right]
\end{equation}
where $D=\sqrt{2{{\gamma }}\lambda -{{\lambda }^{2}}}$, given initial conditions $c(0)=1$ and $\dot{c}(0)=0$~\cite{q-open-syst}. The parameter $D$ distinguishes the two different coupling regimes: an imaginary $D$ corresponds to the weak-coupling regime where ${{\gamma }}<\lambda /2$; a real $D$ corresponds to the strong-coupling regime where ${{\gamma }}>\lambda /2$. It thus makes sense to choose $\alpha=\sqrt{\lambda/\gamma}$ as the universal perturbation parameter. 

Figure~\ref{fig_single} shows approximate solutions of the state population $\rho=\left|c(t)\right|^2$ by the multiple-scale (MS) method compared with the regular perturbation approach that works on the $s$ domain of LT (see Appendix~\ref{appendix_s} for details). In Fig.~\ref{fig_single}, the perturbation solutions of positive orders, LT2 and LT4, as a direct attempt to capture the non-Markovian behavior in the strong-coupling regime, fail to remain bounded. The secular terms in LT2 and LT4 all diverge. In fact, as $\alpha\to0$, the leading term $\alpha^p$ in the memory kernel, Eq.~(\ref{kernel-expansion}), introduces a small boundary layer at $\tilde{t}\sim\alpha^p$, where $c(t)$ may vary so fast that any finite order perturbation term cannot penetrate through the layer and is forced to diverge~\cite{Holmes}. LT(-0) and LT(-2), on the other hand, are approximations derived from inversely perturbing the dynamics, i.e., regular perturbation around $\alpha^{-1}$. LT(-0) shows an exponential decay and should have done well in the weak-coupling regime. When $\alpha^{-1}\gg1$, though, the perturbation series does not converge for the apparent reason.

Instead, MS0 and MS1 are in good agreement with the exact solution. The two closed-form approximations are given by $\left|c^{\left(0\right)}\right|^2$ and $\left|c^{\left(0\right)}+\alpha c^{\left(1\right)}\right|^2$, where
\begin{equation*}
c^{\left(0\right)}(\tilde{T},\tilde{\tau})=f_{+}^{\left(0\right)}(\tilde{\tau})\cos{\left(\frac{G_0^{1/2}}{A_0}\tilde{T}\right)}+f_{-}^{\left(0\right)}(\tilde{\tau})\sin{\left(\frac{G_0^{1/2}}{A_0}\tilde{T}\right)},
\end{equation*}
i.e., $\omega_{\pm}^{\left(0\right)}=\pm G_0^{1/2}/A_0$, given by Eq.~(\ref{integro-differential_ctqw_expand_0}). 
Furthermore, Eq.~(\ref{integro-differential_ctqw_expand_1}) yields
\begin{equation*}
f_{\pm}^{\left(0\right)}(\tilde{\tau})\propto\exp\left(\frac{G_1}{2 G_0 B_0}\tilde{\tau}\right),
\end{equation*}
and the series expansion of $\tilde{G}(\tilde{t}-\tilde{t}')$ yields $G_0=1/2$, $G_1=-1/2$, $G_2=1/4$, $\cdots$. The first order perturbation $c^{\left(1\right)}(\tilde{T},\tilde{\tau})$ follows the exact form of the zeroth order, and $\omega_{\pm}^{\left(1\right)}=\omega_{\pm}^{\left(0\right)}$, while the perturbation correction on $\tilde{T}$ and $\tilde{\tau}$ is given by $A_1/A_0=-1/4$ and $B_1/B_0=0$, respectively. Details of calculation are given in Appendix~\ref{appendix_n1}.

\subsubsection{Non-Markovianity}
The rebound of population shown in Fig.~\ref{fig_single} is one of the features of non-Markovianity. In general, a quantum Markovian evolution is defined by a set of trace-reserving linear maps $\{\mathcal{E}(t,t'),t\ge t'\}$, where $\mathcal{E}(t,t')$ is the propagator for the open system and is required to be \textit{completely positive} under the composition law~\cite{characterize-nMarkov_rhp14}
\begin{equation*}
\mathcal{E}(t,t'')=\mathcal{E}(t,t')\mathcal{E}(t',t''),\qquad t\ge t'\ge t''.
\end{equation*}
This requirement further gives rise to the \textit{Gorini-Kossakowski-Susarshan-Lindblad theorem}~\cite{characterize-nMarkov_rhp14}: a quantum evolution is Markovian if and only if $\mathcal{D}(t)\ge 0$ for all $t$, where $\mathcal{D}(t)$ is named the dissipator in the density matrix formulation~\cite{characterize-nMarkov_rhp14}. Given the series expansion [Eq.~(\ref{tauT})], $\mathcal{D}(t)$ has the following form,
\begin{equation}
\label{dissipator}
\mathcal{D}(T,\tau)=-2\Re\left\{\sum\nolimits_n\alpha^{2n}\left(A_n \alpha\frac{\partial\ln c}{\partial T}+B_n \alpha^{2}\frac{\partial\ln c}{\partial \tau}\right)\right\}.
\end{equation}

Note that the radius of convergence of Eq.~(\ref{dissipator}) about $\alpha$ is given by $\{A_n\}$ and $\{B_n\}$. The Lorentzian reservoir is a special case where $B_n\equiv0$ for $n>1$, indicated by comparing the exact solution [Eq.~(\ref{lorentz_exact})] with $f_{\pm}^{\left(0\right)}(\tilde{\tau})$. In fact, the second part of the sum in Eq.~(\ref{dissipator}) is finite and equal to a positive constant $-\gamma G_1/G_0$ that does not contribute to the non-Markovianity. However, $\Re\{\partial\ln c/\partial T\}$ is not guaranteed to be always negative. Hence the primary time scale $T$ is where non-Markovianity comes from. If $\sum\nolimits_n\alpha^{2n}A_n$ diverges, then the primary time scale must not exist. It is worth noting that expanding $D=\sqrt{2{{\gamma }}\lambda -{{\lambda }^{2}}}$ around $\alpha=0$ yields
\begin{equation*}
D \simeq \sqrt{2} \gamma \left(\alpha-\alpha^3/4+\cdots\right),
\end{equation*}
of which the coefficients coincide with $\{A_n\}$. Meanwhile, the system is non-Markovian if and only if $D$ is real. Therefore, we are convinced that the radius of convergence determined by the sequence $\{A_n\}$ is the dividing line between Markovian regime and non-Markovian regime, a result accompanied by the emergence of another independent time scale.

Another feature of non-Markovianity related to time scales is the steep decrease of quantum speed limit in the strong-coupling regime~\cite{QSL_mwg15}. It is found that the evolution time between two orthogonal pure or mixed states may not be unique if the evolution is non-Markovian~\cite{QSL_mwg15}. In Eq.~(\ref{dissipator}), there are infinite numbers of singularities along the time axis, set by $c(t)=0$. They correspond to all possibilities of the evolution time between the two eigenstates of $H_{\mathcal{G}}$. The first singularity is approximately at ${{\hat{t}}_{\text{MS0}}}=\pi (2\gamma\lambda)^{-1/2}$, or ${{\hat{t}}_{\text{MS1}}}=(\gamma \lambda /2)^{-1/2}(1- \lambda/4\gamma)^{-1}{{\arccos }}\sqrt{\lambda /\left( 2\gamma+\lambda  \right)}$, derived by MS0 and MS1, respectively. Their relative errors are only to the order of $\alpha$ and $\alpha^3$, compared to the exact result $\hat{t}=(2/D)[\pi - \arctan(D/\lambda)]$ which is known as the \emph{minimal evolution time}~\cite{QSL_mwg15} between the two orthogonal eigenstates. 

\subsection{General reservoirs}
\label{section_reservoir}
It is known that reservoir engineering helps producing squeezed states~\cite{reservoir-engineer_klkcllmnh15} or generating non-Markovianity~\cite{reservoir-engineer_llhlglbp11}. A general and well-behaved approximation method is potentially useful for this practical purpose. Here, we apply the MS perturbation method on more general reservoirs and investigate how the goodness of approximation is related to the form of memory kernel.

As shown in Figs.~\ref{fig_reservoirs}\subref{fig_reservoirs_a}-\ref{fig_reservoirs}\subref{fig_reservoirs_c}, we choose three different simple but nontrivial memory kernels, where there are still two parameters, $\gamma$ and $\lambda$, by which $\alpha\sim(\lambda/\gamma)^{1/2}$ is to be fully constructed. Their forms are similar to the Lorentzian reservoir, but their series expansions in terms of $\{G_n\}$ are clearly different. We define the characteristic times, $T_c=\gamma^{-1}G_0^{-1/2}$ and $\tau_c=-\gamma^{-1}G_0/G_1$, for the primary and auxiliary time scales, respectively. 
Their relations with the memory kernels [Figs.~\ref{fig_reservoirs}\subref{fig_reservoirs_a}-\ref{fig_reservoirs}\subref{fig_reservoirs_c}] imply that the global behavior of $c(t)$ is determined only locally by $G(t-t')$ at $t\to t'$ if $G(t)$ is holomorphic. Thus the goodness of approximation depends on how well the series expansion of $G(t)$ behaves in the local range.

To be more specific, in Figs.~\ref{fig_reservoirs}\subref{fig_reservoirs_a}~and~\ref{fig_reservoirs}\subref{fig_reservoirs_d}, the memory kernel is of the form of a Gauss error function, $G(t-t')=(\pi^{3/2}\gamma\lambda/\sqrt{2}) \text{erfc}(\lambda\left|t-t'\right|/\sqrt{2})$. As expected, the approximate solutions MS0 and MS1 are good enough, because $\text{erfc}(x)\sim O(x^{-1}\exp(-x^2))$ descends fast enough when $x\to\infty$, making $G(t-t')$ more dominated by local behavior. In Figs.~\ref{fig_reservoirs}\subref{fig_reservoirs_b}~and~\ref{fig_reservoirs}\subref{fig_reservoirs_e}, $G(t-t')=\gamma\lambda/(1+\lambda\left|t-t'\right|)$. A descending speed of $O(x^{-1})$ makes $G(t-t')$ less dominated by local behavior and thus implies the approximation to be less accurate. Note that $O(x^{-1})$ also induces logarithmic singularity in the corresponding spectral density. In Figs.~\ref{fig_reservoirs}\subref{fig_reservoirs_c}~and~\ref{fig_reservoirs}\subref{fig_reservoirs_f}, it is a special case that we have $\tau_c\to\infty$ implied by the Gaussian memory kernel, $G(t-t')=\sqrt{2\pi}\gamma\lambda\exp(-\lambda^2\left|t-t'\right|^2/2)$. The auxiliary scale $\tau$ collapses to an infinitesimal point and thus becomes trivial. Given that there is only one attractor at $c(\infty)=0$~\cite{structured-optical-reservoirs_lnnb00}, we believe another time scale that corresponds to decay behavior (which, even though existing, must be comparably small) should be hidden in the system with a complex nonlinear manner. This raises the question of how to identify all time scales in a general reservoir, which is theoretically important yet a nontrivial task.

\section{Regular networks}
\label{section_regular-netw}
In this section, we investigate how network topology determines the propagation of an open-system walker. We study regular networks, e.g., complete networks, star networks, rings, and square lattices, where repeating patterns exist in their network topologies. All results given hereafter are calculated by MS0, i.e., only to the zeroth order. We omit the superscript, $\omega^{(0)}\equiv \omega$, and let $A_0=B_0=1$, w.l.o.g., for simplification. Only the two leading terms $G_0$ and $G_1$ from $\{G_n\}$ are used, and we set them to be $G_0=1/2$ and $G_1=-1/2$ which coincide with those of the Lorentzian reservoir. The reservoir itself, however, does not have to be of the exact same type.

\subsection{Binary quantum walk}
The simplest quantum walk is a binary system, i.e., $N=2$. The walker simply chooses between ``Yes'' and ``No'', the two nodes of $\mathcal{G}$. Without any reservoir, the binary system is a simple harmonic oscillator, with a 
frequency of $|\omega_0^{\pm}|=\alpha^{-1}A_{12}\sim T^{-1}$. Here, we assume that there are two eigenfrequencies, $\omega_0^{+}$ and $\omega_0^{-}$, as we will see how the degeneracy is broken by introducing a reservoir.

\begin{figure}[t]
	\includegraphics[width=8.6cm]{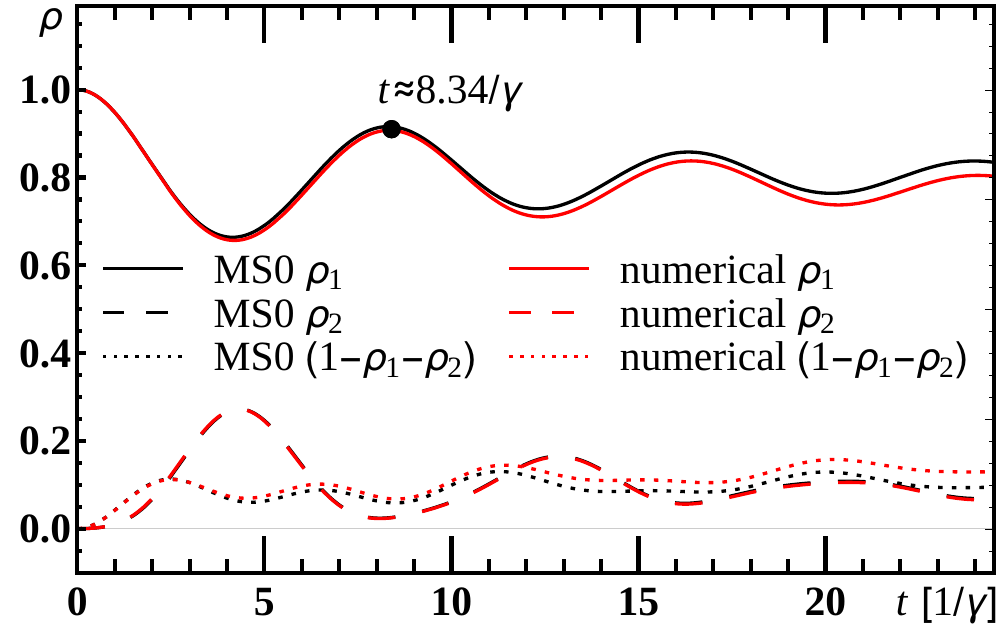}
	\caption{\label{fig_double}(Color online) State populations of a binary system in a Lorentzian reservoir. $\lambda =\gamma/10$, $w_1=1$, $w_2=3$, and $A_{12}=A_{21}=0.08\gamma$.\hfill \hfill}
\end{figure}

Figure~\ref{fig_double} shows how the state populations $\rho_1=\left|c_1\right|^2$ and $\rho_2=\left|c_2\right|^2$ change over time and how they are compared with the numerical solution in a Lorentzian reservoir. Besides the goodness of approximation, we note that the oscillation frequency between $\rho_1$ and $\rho_2$ is indeed significantly increased, as $(8.34/\gamma)^{-1}\pi\approx0.377\gamma\gg0.08\gamma$, a consequence of the non-Markovianity-assisted steep decrease of quantum speed limit~\cite{QSL_mwg15}. By solving Eq.~(\ref{integro-differential_ctqw_expand_0}), we actually find three eigenfrequencies, $\omega_0^{+}\approx-2.32\gamma$, $\omega_0^{-}\approx2.17\gamma$, and $|\omega_{\infty}^{\pm}|\approx0.151\gamma$, which satisfy $\omega_0^{+}+\omega_0^{-}+|\omega_{\infty}^{\pm}|=0$. The magnitudes of $\omega_0^{+}$ and $\omega_0^{-}$ are larger than $\alpha^{-1}A_{12}\approx0.253\gamma$, while $|\omega_{\infty}^{\pm}|$ is smaller. We use the notations $\omega_0^{\pm}$ and $\omega_{\infty}^{\pm}$ for the sake of consistency with the following context where we will see that at most four eigenfrequencies exist, regardless of the coupling ratio of individual nodes.

One way to implement a binary quantum walk in cavity QED is to produce a dipole-dipole interaction~\cite{dipole-dipole-interaction_lzg09} between two atoms. Oscillation between the two upper levels of the atoms is regarded as the quantum walk between two nodes. Our perturbation method provides a convenient way to study exchange of states and its speed limit in relevant dipole-dipole interaction experiments.

\begin{figure*}[t!]
	\begin{centering}			
		\subfloat[ ]{
			\label{fig_complete-netw_3_w}
			\includegraphics[width=5.7cm]{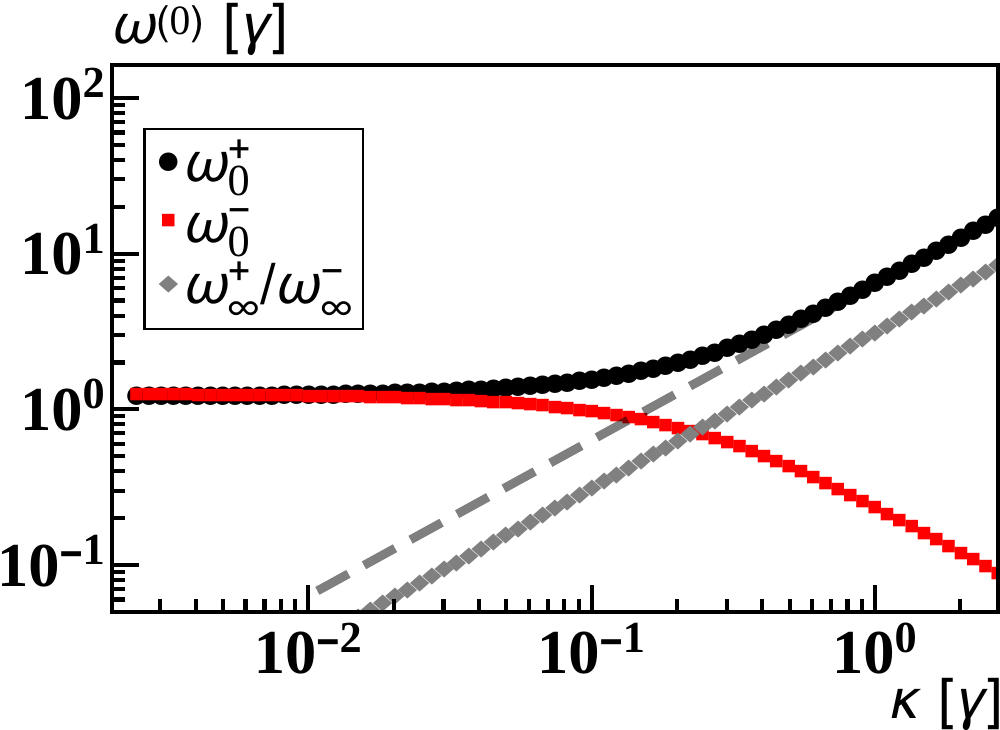}			
		}\subfloat[ ]{
			\label{fig_complete-netw_100_w}
			\includegraphics[width=5.7cm]{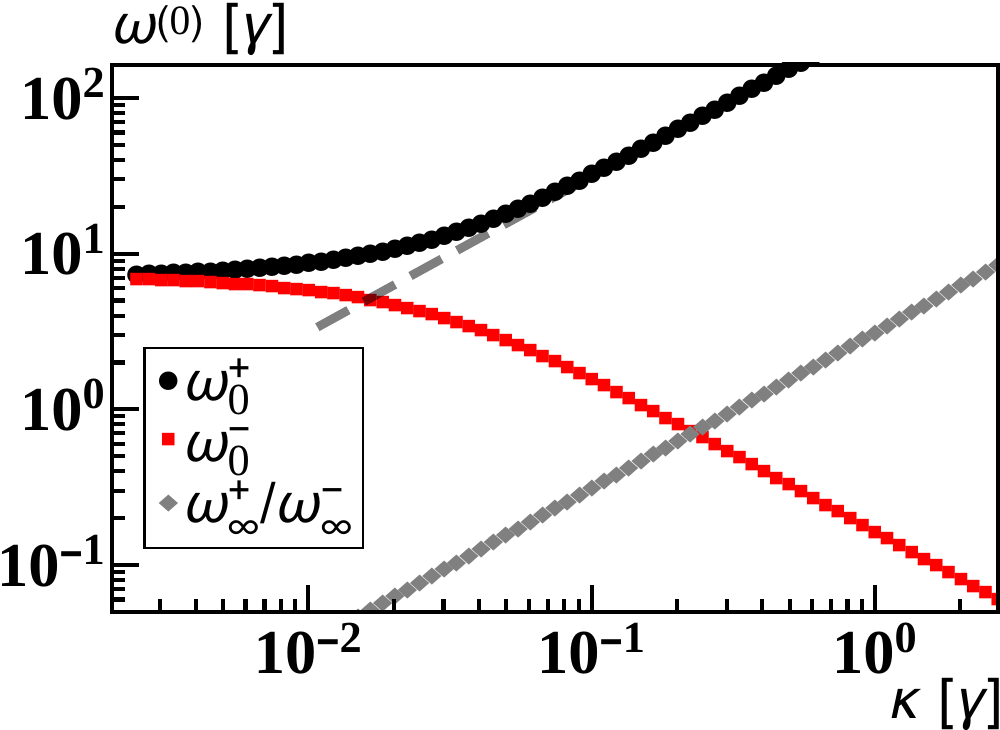}				
		}\subfloat[ ]{
			\label{fig_complete-netw_100_rand_w}
			\includegraphics[width=5.7cm]{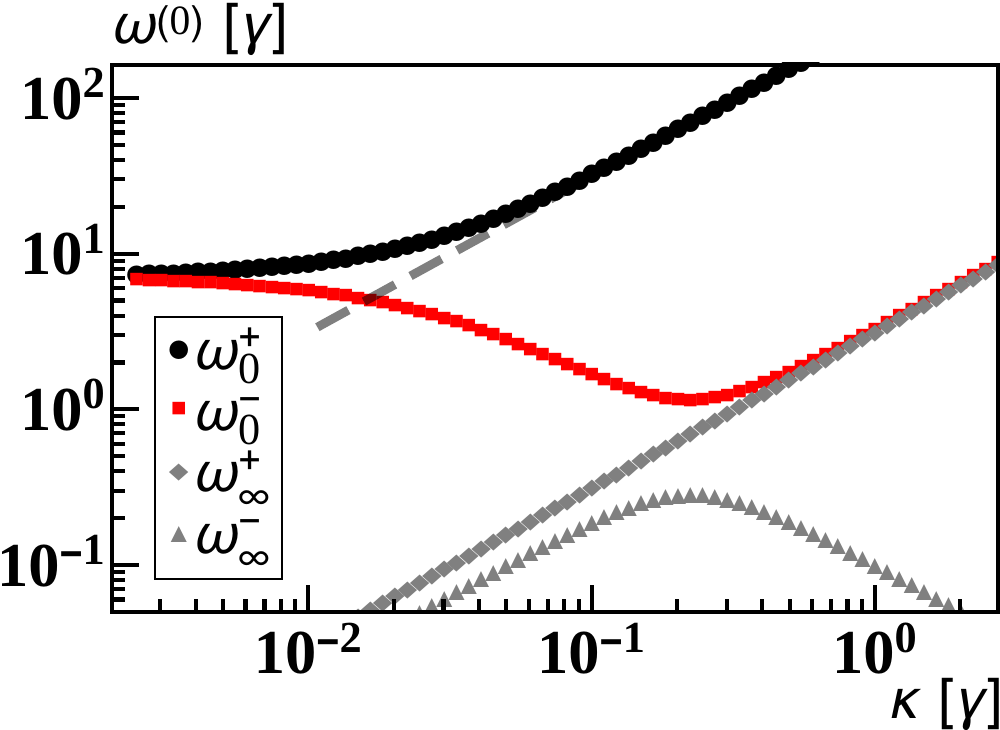}				
		}
		
		\subfloat[ ]{
			\label{fig_complete-netw_3_g}
			\includegraphics[width=5.7cm]{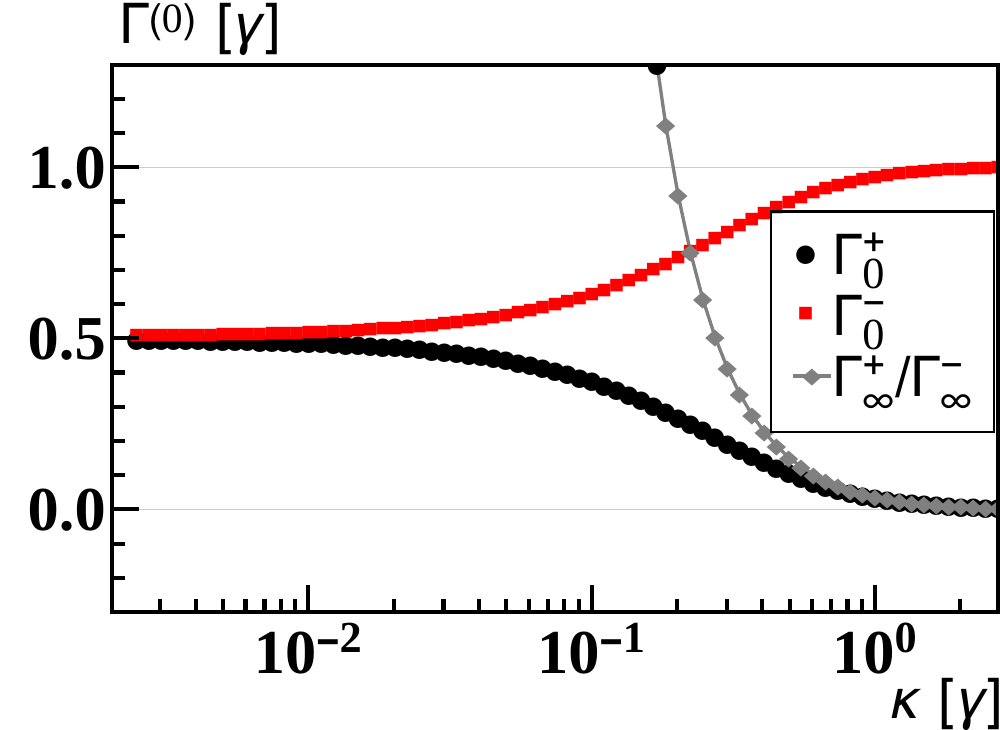}			
		}\subfloat[ ]{
			\label{fig_complete-netw_100_g}
			\includegraphics[width=5.7cm]{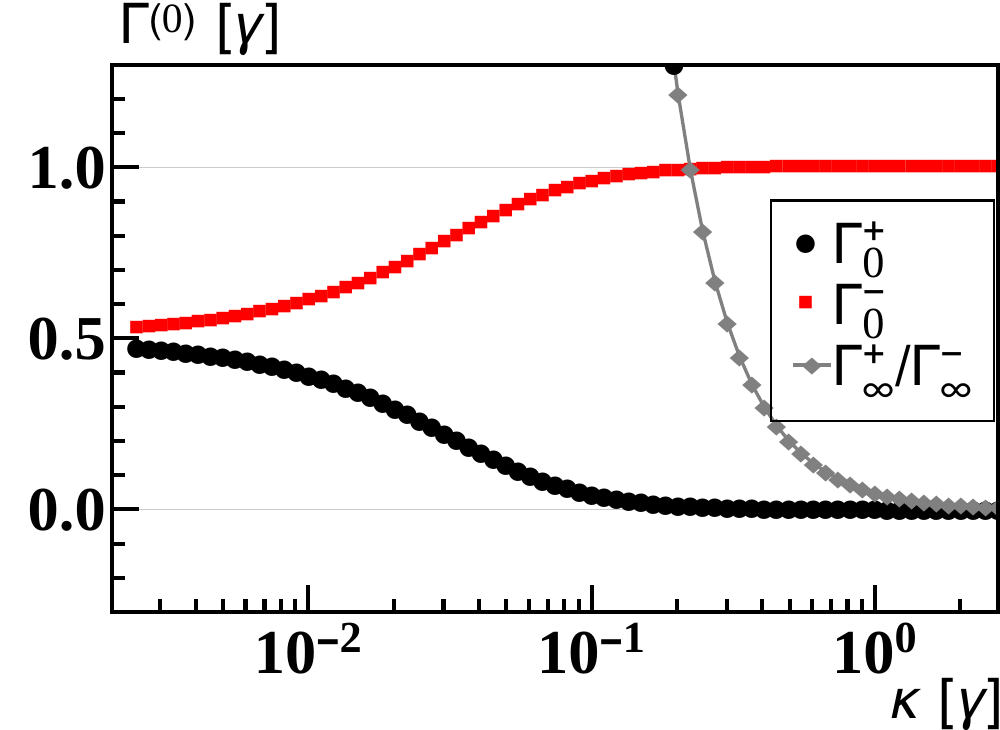}			
		}\subfloat[ ]{
			\label{fig_complete-netw_100_rand_g}
			\includegraphics[width=5.7cm]{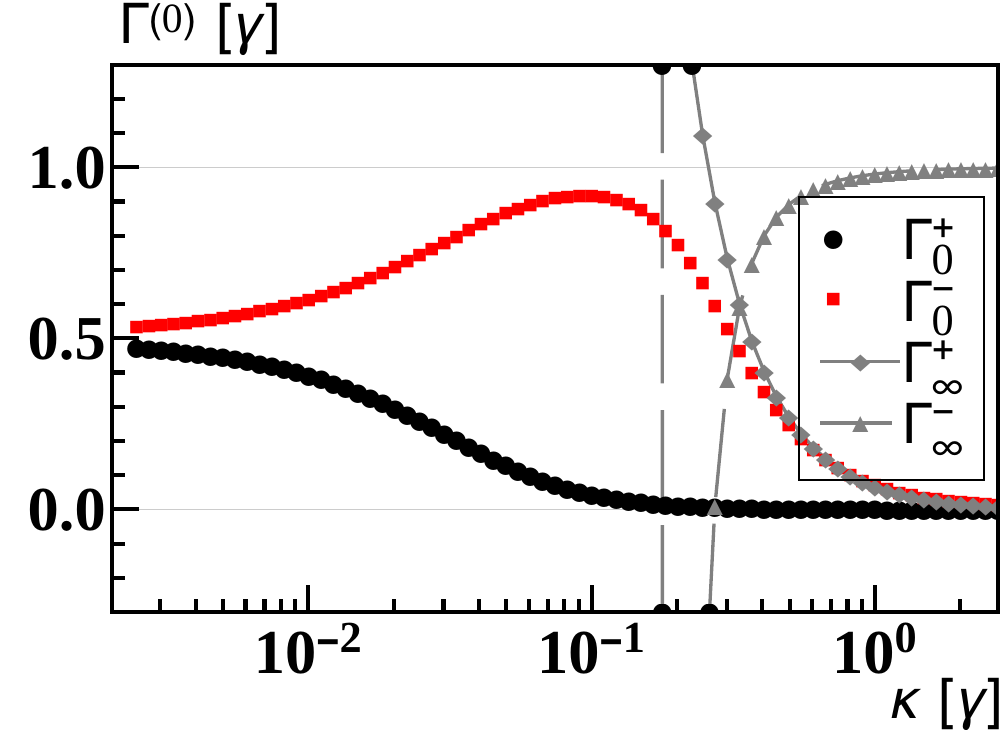}			
		}	
	\end{centering}
	
	\caption{\label{fig_complete-netw} (Color online) Quantum walk on a complete network $\mathcal{G}$, enclosed in a reservoir with its memory kernel characterized by $G_0=1/2$ and $G_1=-1/2$, the coupling ratio $w_i=1$ being the same for each node $i$. Here shown are the eigenfrequencies, $\omega_0^{\pm}$ and $\omega_{\infty}^{\pm}$ (their magnitudes) [by Eq.~(\ref{integro-differential_ctqw_expand_0})], and their corresponding decay rates, $\Gamma_0^{\pm}$ and $\Gamma_{\infty}^{\pm}$ [by Eq.~(\ref{integro-differential_ctqw_expand_1})], with respect to $\kappa$, the weight on each link of $\mathcal{G}$. (a),~(d): $N=3$. (b),~(e): $N=100$. (c),~(f): $N=100$, while $w_i$ is no longer constant but follows a slightly random uniform distribution, $\mathbf{w}\sim\mathcal{U}^{N}(0.9,1.1)$.\hfill \hfill}
\end{figure*}

\subsection{Complete network}
When $N>2$, network topology starts to have an effect on the system dynamics. We let $\mathcal{G}$ to be a complete network, so that
\begin{equation*}
A_{ij}=\left\{
\begin{array}{ll}
\kappa &\quad i\neq j \\
0 & \quad i=j
\end{array}
\right..
\end{equation*}
Here, $\kappa$ is the link weight which has a dimension of $\gamma$. Figure~\ref{fig_complete-netw} shows the magnitudes of $\omega_0^{\pm}$ and $\omega_{\infty}^{\pm}$ [by Eq.~(\ref{integro-differential_ctqw_expand_0})] and their corresponding decay rates $\Gamma_0^{\pm}$ and $\Gamma_{\infty}^{\pm}$ [by Eq.~(\ref{integro-differential_ctqw_expand_1})] as functions of $\kappa$. It is worth noting that while $\Gamma_0^{+}$ and $\Gamma_0^{-}$ are bounded, $\Gamma_{\infty}^{\pm}$ diverge when $\kappa\to0$, which is unphysical. To derive a physical solution, we are forced to reassign $\Gamma_{\infty}^{\pm}$ to another universal and trivial solution of Eq.~(\ref{integro-differential_ctqw_expand_1}) which we ignored first, namely, $\Gamma_{\infty}^{\pm}\equiv0$. The deeper implication of abandoning the diverging $\Gamma_{\infty}^{\pm}$ is, as first revealed in Fig.~\ref{fig_reservoirs}\subref{fig_reservoirs_f}, that there should be another global and complex time scale(s) which is responsible for long-term decay. Therefore, the approximate solution of the quantum walker state reads
\begin{eqnarray*}
\mathbf{c}({T},{\tau})&\approx&\mathbf{C}_{\infty}^{+}e^{-i \omega_{\infty}^{+} {T}}+\mathbf{C}_{\infty}^{-}e^{-i \omega_{\infty}^{-} {T}}\nonumber\\
&&+\mathbf{C}_0^{+}e^{-\Gamma_0^{+}{\tau}}e^{-i \omega_0^{+} {T}}
+\mathbf{C}_0^{-}e^{-\Gamma_0^{-}{\tau}}e^{-i \omega_0^{-} {T}}.
\end{eqnarray*}
In fact, by numerical solution, we find that those terms related to the long-term frequencies $\omega_{\infty}^{\pm}$ indeed decay much slower, the magnitudes of which can be approximated as constants $\mathbf{C}_{\infty}^{\pm}$ in the short run.
 
As $\kappa\to\infty$, Figs.~\ref{fig_complete-netw}\subref{fig_complete-netw_3_w}-\ref{fig_complete-netw}\subref{fig_complete-netw_100_rand_w} together show that
$\omega_{0}^{+}$ is asymptotically close to $(N-1)\kappa/\alpha$ (dashed line), which is the intrinsic mode frequency of a complete network. At the same time, $\Gamma_{0}^{+}\to0$ guarantees that the intrinsic mode frequency is secular in the $\kappa\to\infty$ limit, and thus we recover a CTQW as a closed system. 

When $\kappa\to0$, $\omega_{0}^{+}$ and $\omega_{0}^{-}$ approach a constant, $N^{1/2}G_0^{1/2}$, which is the oscillation frequency induced by the reservoir. Since $\omega_{0}^{\pm}$ are directly relevant to non-Markovianity, we deduce that the existence of short-term frequencies is responsible for improvement of quantum speed limit in the non-Markovian regime. Such improvement, however, is only temporary and is controlled by the short-term decay rates $\Gamma_{0}^{\pm}\to-(2G_0)^{-1}G_1$, because in the long run only long-term frequencies $|\omega_{\infty}^{\pm}|$ persist which are smaller than $(N-1)\kappa/\alpha$, indicating a slowdown of dynamics.
 
We also see that the degeneracy between $\omega_{\infty}^{+}$ and $\omega_{\infty}^{-}$, as well as $\Gamma_{\infty}^{+}$ and $\Gamma_{\infty}^{-}$, is broken by introducing small fluctuation on $\mathbf{w}$, as shown in Figs.~\ref{fig_complete-netw}\subref{fig_complete-netw_100_rand_w}~and~\ref{fig_complete-netw}\subref{fig_complete-netw_100_rand_g}. We expected more eigenfrequencies to be split from the eigenspectrum of Eq.~(\ref{integro-differential_ctqw_expand_0}), but only see four frequencies at most, no matter how large $N$ is. The hidden degeneracies are thus maintained by the topological symmetry of $\mathcal{G}$. Note that although $\mathbf{w}$ is randomly realized in Figs.~\ref{fig_complete-netw}\subref{fig_complete-netw_100_rand_w}~and~\ref{fig_complete-netw}\subref{fig_complete-netw_100_rand_g}, different realizations of $\mathbf{w}$ indeed yield similar results, since the ensemble average of different realizations actually converges to the one-shot observation when $N$ becomes large, which is guaranteed by the assumed ergodicity of random matrix theory. 

Finally, we note that in spite of its speed-up effect on quantum walks, open-system dynamics usually forbids perfect state transfer~\cite{perfect-state-transfer_k10} in the studied system. Hence, introducing a reservoir brings both advantages and disadvantages, between which the trade-off needs to be taken care of by delicate reservoir engineering.

\begin{figure*}[t!]
		\begin{centering}			
		\subfloat[ ]{
			\label{fig_star-netw_3_w}
			\includegraphics[width=5.7cm]{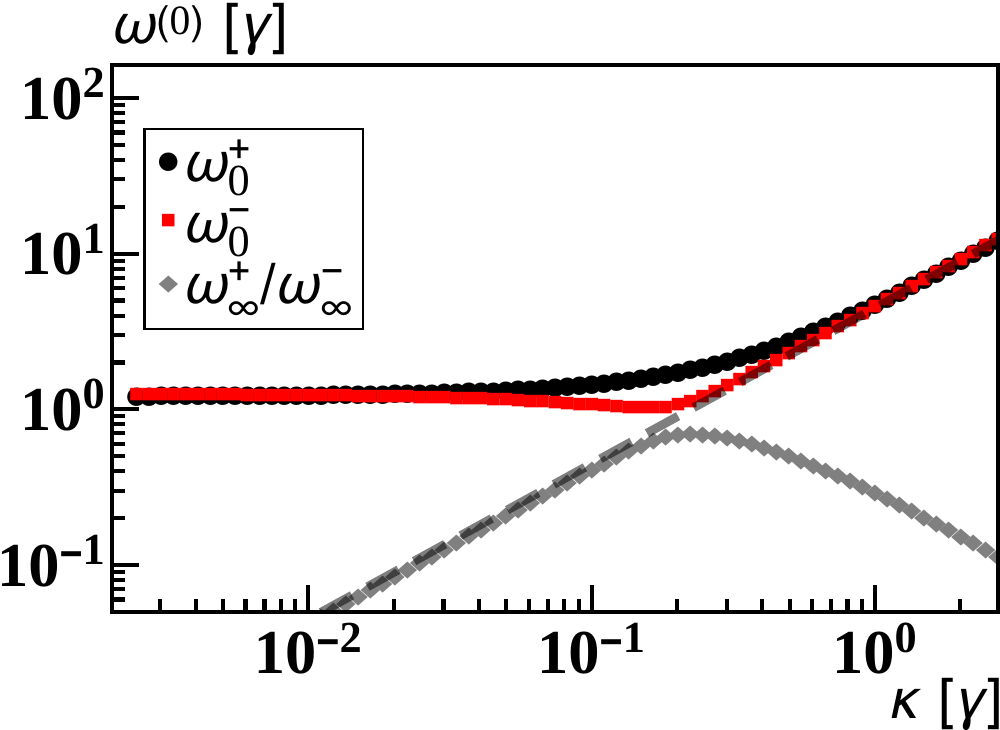}			
		}\subfloat[ ]{
			\label{fig_star-netw_100_w}
			\includegraphics[width=5.7cm]{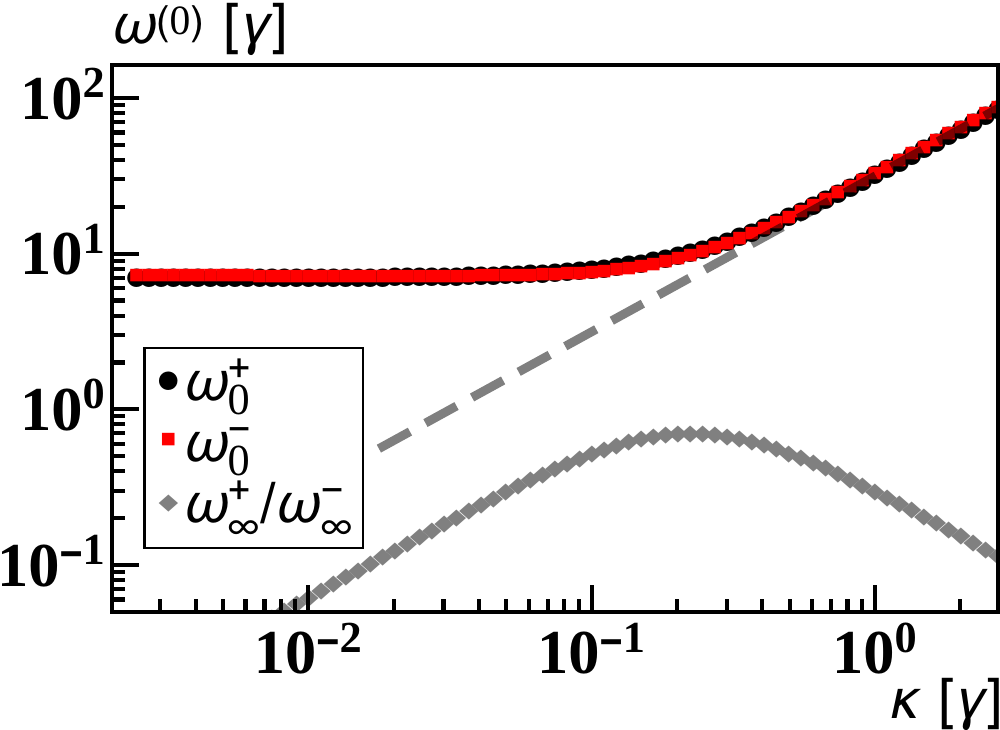}				
		}\subfloat[ ]{
			\label{fig_star-netw_100_rand_w}
			\includegraphics[width=5.7cm]{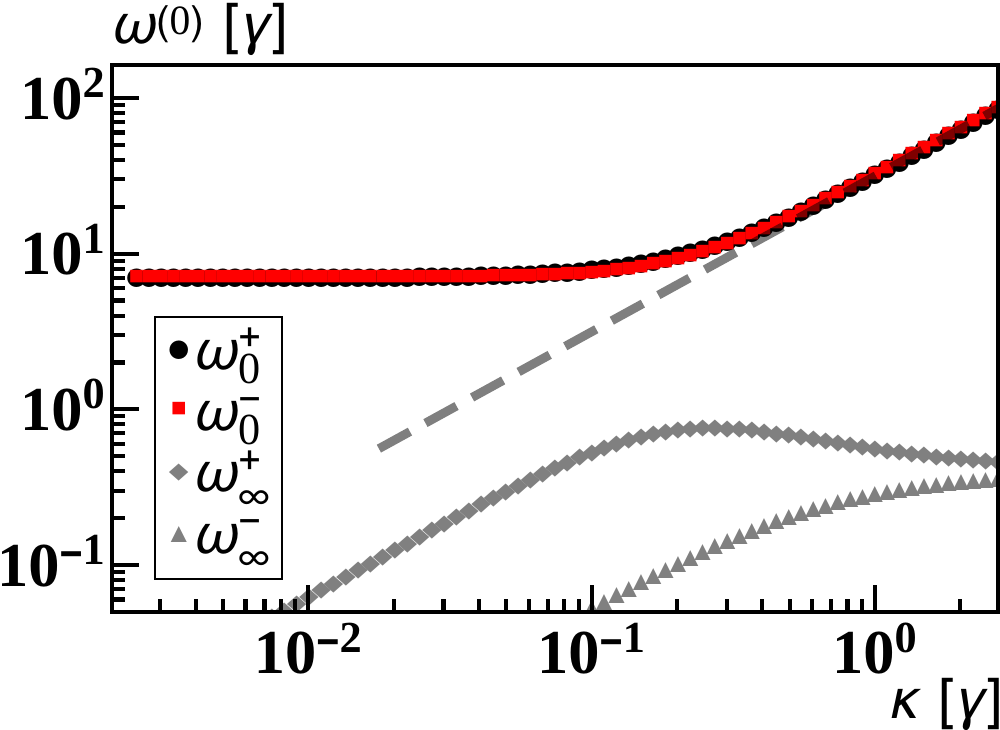}				
		}
		
		\subfloat[ ]{
			\label{fig_star-netw_3_g}
			\includegraphics[width=5.7cm]{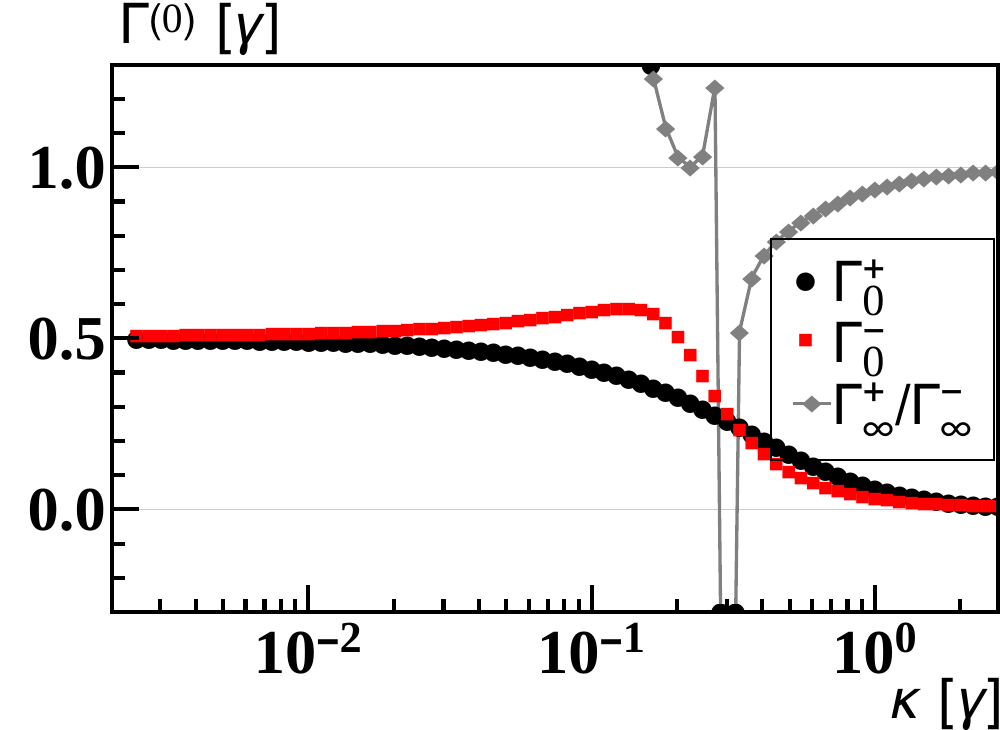}			
		}\subfloat[ ]{
			\label{fig_star-netw_100_g}
			\includegraphics[width=5.7cm]{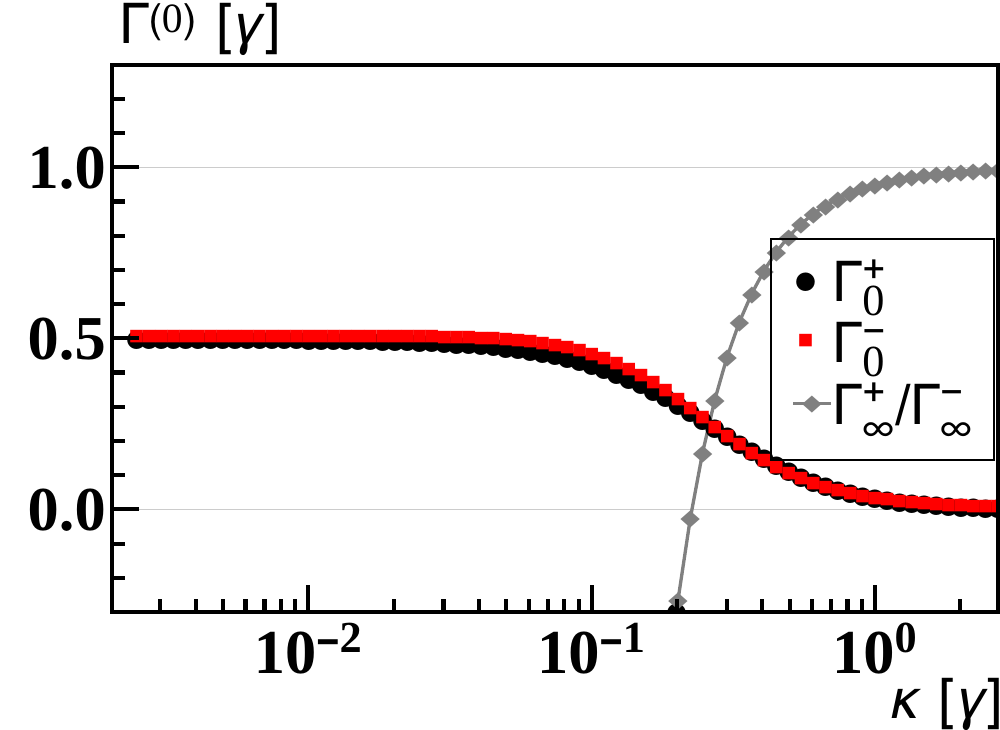}			
		}\subfloat[ ]{
			\label{fig_star-netw_100_rand_g}
			\includegraphics[width=5.7cm]{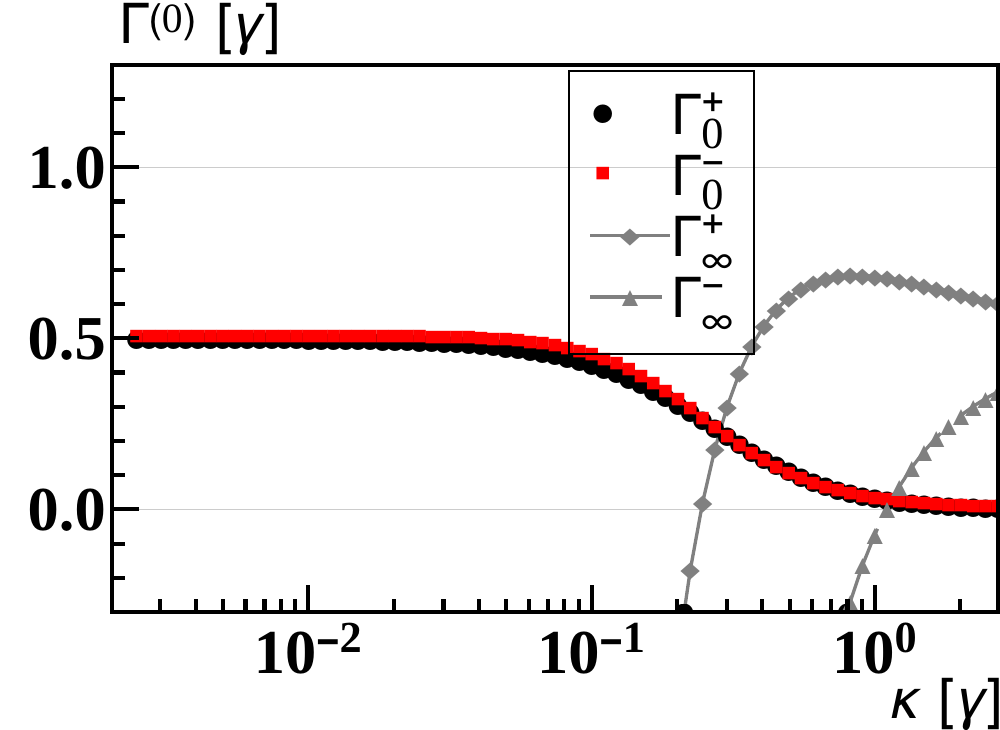}			
		}	
	\end{centering}

\caption{\label{fig_star-netw} (Color online) Quantum walk on a star network. Configurations are the same as in Fig.~\ref{fig_complete-netw}.\hfill \hfill	
}
\end{figure*}

\subsection{Star network}
Instead, we let $\mathcal{G}$ to be a star network, i.e.,
\begin{equation*}
A_{ij}=\left\{
\begin{array}{ll}
\kappa &\quad i=1,j\neq1 \\
0 & \quad \text{otherwise}
\end{array}
\right..
\end{equation*}
All results are presented in Fig.~\ref{fig_star-netw} which shares a lot of similarity with Fig.~\ref{fig_complete-netw}. Note however that the difference between a star network and a complete network is fundamental, as there is always proximate degeneracy between $\omega_{0}^{+}$ and $\omega_{0}^{-}$, as well as $\Gamma_{0}^{+}$ and $\Gamma_{0}^{-}$ (Fig.~\ref{fig_star-netw}). In particular, comparing Figs.~\ref{fig_star-netw}\subref{fig_star-netw_3_w}~and~\ref{fig_star-netw}\subref{fig_star-netw_3_g} with Figs.~\ref{fig_complete-netw}\subref{fig_complete-netw_3_w}~and~\ref{fig_complete-netw}\subref{fig_complete-netw_3_g} demonstrates how different network topologies affect the emergence of a new eigenvalue in their corresponding spectrums when $N=3$. Also, as shown in Figs.~\ref{fig_star-netw}\subref{fig_star-netw_3_w}-\ref{fig_star-netw}\subref{fig_star-netw_100_rand_w}, the intrinsic mode frequency of a star network is $\sqrt{N-1}\kappa/\alpha$ (dashed line), which is smaller than that of a complete network. $|\omega_{\infty}^{\pm}|\le\sqrt{N-1}\kappa/\alpha$ still holds true. Nevertheless, we see that both $\Gamma_{0}^{\pm}\to0$ as $\kappa\to\infty$, revealing that the short-term decay rates are lower when $\kappa$ is not too small---a potential advantage to prolong the speed-up of quantum walks.

\begin{figure}[t]
	\begin{centering}			
		\subfloat[ ]{
			\label{fig_ring-netw_100_rand_w}
			\includegraphics[width=5.7cm]{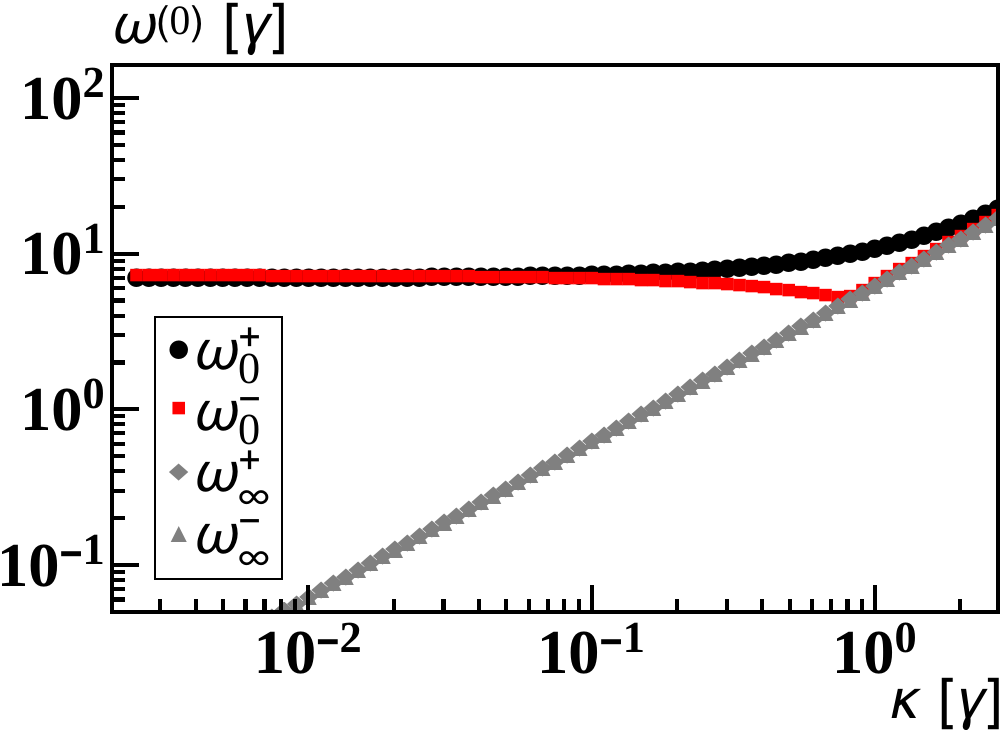}				
		}
	
		\subfloat[ ]{
			\label{fig_ring-netw_100_rand_g}
			\includegraphics[width=5.7cm]{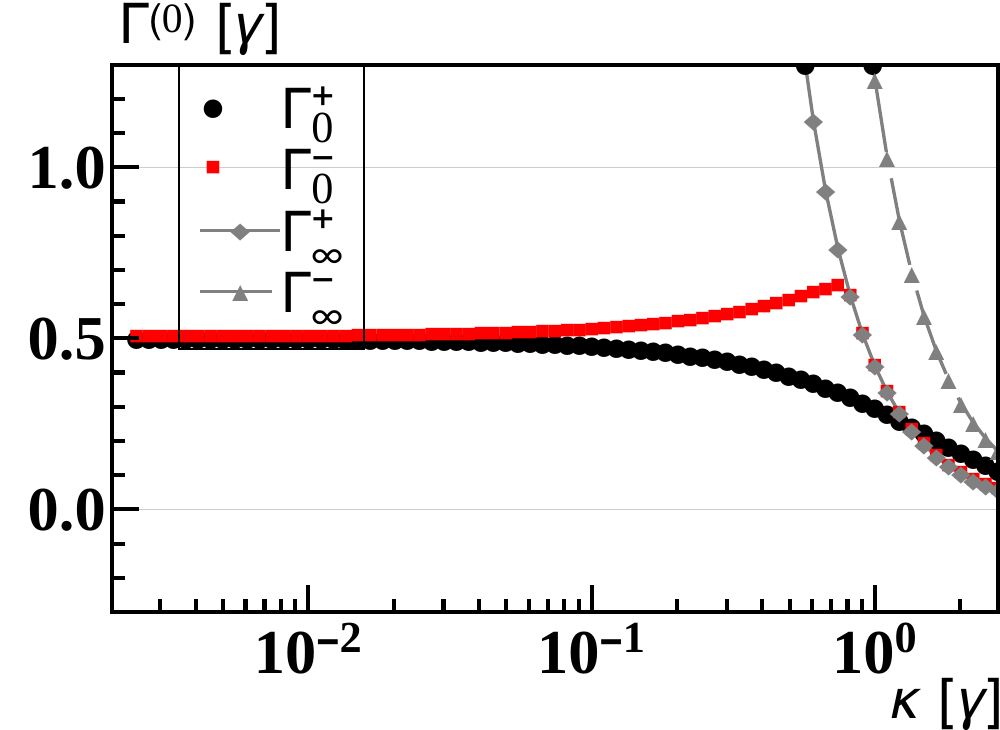}			
	}	
\end{centering}

\caption{\label{fig_ring-netw} (Color online) Quantum walk on a ring. Configurations are the same as in Figs.~\ref{fig_complete-netw}(c)~and~\ref{fig_complete-netw}(f). \hfill \hfill}
\end{figure}

\begin{figure}[t]
	\begin{centering}			
		\subfloat[ ]{
			\label{fig_square-netw_100_rand_w}
			\includegraphics[width=5.7cm]{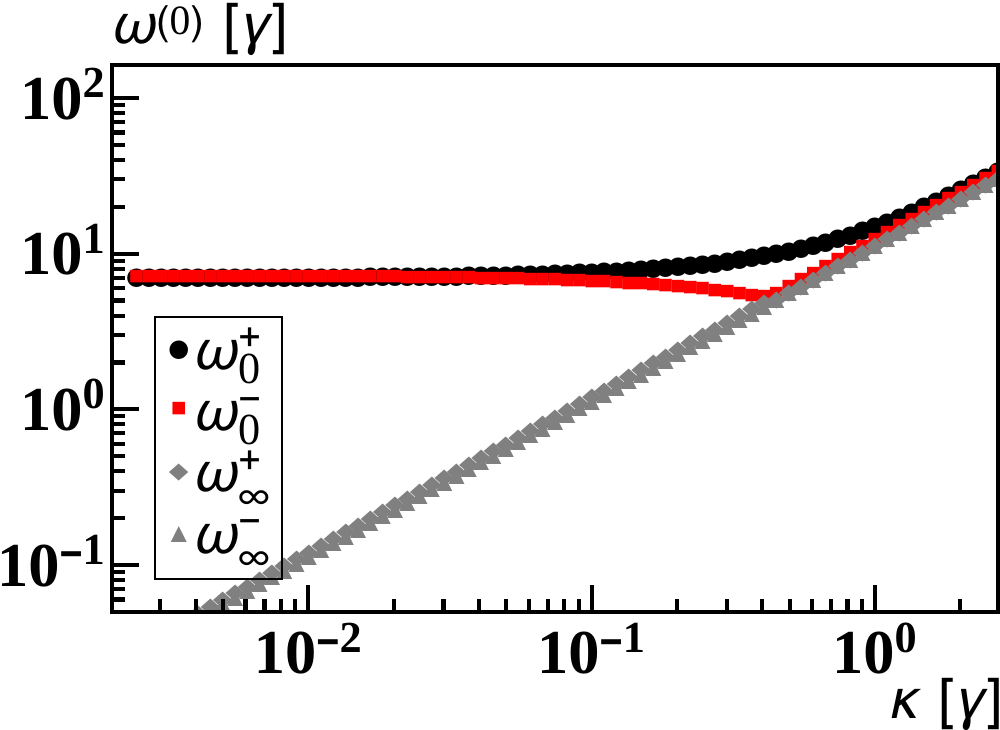}				
		}
	
		\subfloat[ ]{
			\label{fig_square-netw_100_rand_g}
			\includegraphics[width=5.7cm]{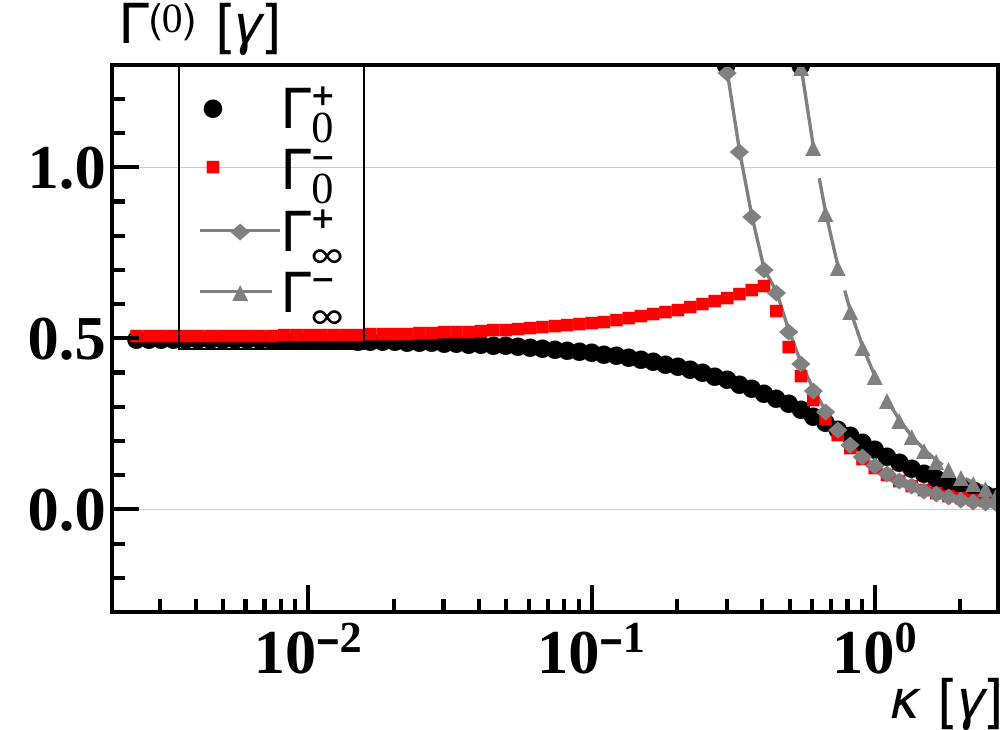}			
		}	
	\end{centering}

\caption{\label{fig_square-netw} (Color online) Quantum walk on a square lattice. Configurations are the same as in Figs.~\ref{fig_complete-netw}(c)~and~\ref{fig_complete-netw}(f). \hfill \hfill}
\end{figure}

\subsection{Ring}
If $\mathcal{G}$ is a ring, we have
\begin{equation*}
	A_{ij}=\left\{
	\begin{array}{ll}
		\kappa &\quad i-1=j \text{ mod } N \text{, or } j-1=i \text{ mod } N\\
		0 & \quad \text{otherwise}
	\end{array}
	\right..
\end{equation*}
In Fig.~\ref{fig_ring-netw}, it is reconfirmed that at most four eigenfrequencies exist in $\mathcal{G}$. The unphysical decay rates $\Gamma_{\infty}^{\pm}$ still diverge as expected. It is also shown that the degeneracy between $\omega_{\infty}^{+}$ and $\omega_{\infty}^{-}$ is barely broken even with fluctuation on $\mathbf{w}$ turned on. This is because the diameter of a ring is $\sim N/2$, much larger than the diameter of a complete network or a star network which is only $1$ or $2$. Thus the fluctuation on $\mathbf{w}$ only has a local impact which is averaged at long distance. It is also interesting to see that $\omega_{0}^{\pm}$ and $\omega_{\infty}^{\pm}$ all approach $2\kappa/\alpha$ asymptotically as $\kappa\to\infty$. There are no other eigenfrequencies.

\subsection{Square lattice}
When $\mathcal{G}$ is a $\sqrt{N}\times\sqrt{N}$ square lattice, the behaviors of eigenfrequencies and decay rates of $\mathcal{G}$ (Fig.~\ref{fig_square-netw}) are similar to those of a ring. Comparing Fig.~\ref{fig_square-netw}\subref{fig_square-netw_100_rand_w} with Fig.~\ref{fig_ring-netw}\subref{fig_ring-netw_100_rand_w}, we see that instead $\omega_{0}^{\pm}$ and $\omega_{\infty}^{\pm}$ approach $4\kappa/\alpha$ asymptotically as $\kappa\to\infty$. Since a ring is a $1$-D system and a square lattice is $2$-D, we conclude that for any finite dimensional system, the spectrum of Eq.~(\ref{integro-differential_ctqw_expand_0}) should be simpler than infinite dimensional networks.

\section{Conclusion}
\label{section_conclus}
To summarize, we present a multiple-scale perturbation method that works on integro-differential equations in the form of Eq.~(\ref{integro-differential}), which can be used to unravel the functional importance hidden in the memory kernel and its related complex dynamics. The multiple-scale perturbation method helps to find closed-form approximate solutions that cannot be derived from regular integral transforms, providing a controllable precision that meets practical needs. 

In particular, we study its application to a continuous-time quantum walk on some network $\mathcal{G}$ enclosed by a general non-Markovian reservoir $\mathcal{E}$. Such a composite system can be regarded as a quantum walk between possible ``target'' sites ($\mathcal{G}$) and ``error'' sites ($\mathcal{E}$) if viewed as a quantum error correction algorithm. We propose two physically-important time scales, a primary time scale $T$ and an auxiliary time scale $\tau$, both existing in the strong-coupling regime where $\mathcal{G}$ and $\mathcal{E}$ are strongly coupled. Compared to the failure of ordinary perturbations supported by Laplace transform, the multiple-scale method shows sufficient accuracy which should be determined by how fast the memory kernel converges locally. The emergence of a new time scale, as the coupling goes stronger, is closely related to the emergence of non-Markovianity. Next, we investigate the eigenfrequencies and their corresponding decay rates of quantum walks on different regular networks. The speed-up of non-Markovian dynamics is confirmed, which however does not exist in the long run, when the two short-term fast frequencies become negligible and only the two long-term slow frequencies persist, which are smaller than the intrinsic mode frequency (the frequency when no reservoir exists). In addition, the behaviors of quantum walks on rings and square lattices are rather simpler than those on complete networks and star networks, because of the limit from dimensionality.

Further studies on other issues of continuous-time quantum walks should be carried on, through which we hope that the two pertinent practical areas---reservoir engineering and quantum search algorithm design---which for now are the most focused can be substantially developed soon.
\newline

\begin{acknowledgements}
We thank Bin Luo, Jun-Hong An, and Chengjun Wu for fruitful discussions. Y.L. acknowledges the valuable comments from H. M. Wiseman and W. T. Strunz. X.M. and H.E.S. acknowledge the assistance from J. Morrow. Y.L., J.-W.Z., and H.G. are supported by National Natural Science Foundation of China (NSFC Grant Nos.~61571018, 61531003, 91436210) and National Key Research and Development Program (NKRDP). X.M. and H.E.S. are supported by DTRA Grant HDTRA1-14-1-0017.
\end{acknowledgements}

\appendix
\section{Regular perturbation method on the $s$ domain of Laplace transform}
\label{appendix_s}
For a $N=1$ system in a Lorentzian reservoir at resonance ($\Delta=0$), combining Eqs.~(\ref{integro-differential_ctqw})~and~(\ref{Gexp}) yields
\begin{equation}
\label{lorentz_exact_s}
s c(s)-1=-\frac{1}{2}\frac{\alpha^2}{s+\alpha^2}c(s),
\end{equation}
where $c(s)=\int_{0}^{\infty}{d\tilde{t} c(\tilde{t})e^{-s\tilde{t}}}$ is the Laplace transform of $c(\tilde{t})$. The initial conditions at $t=0$ are chosen $c(0)=1$ and $\dot{c}(0)=0$. Note that Eq.~(\ref{lorentz_exact_s}) has a simple closed-form solution, i.e., Eq.~(\ref{lorentz_exact}), thanks to the exponential form of the memory kernel [Eq.~(\ref{Gexp})]. $c(s)$ is further expanded around $\alpha$, $c(s)=\sum\nolimits_{n}\alpha^{2n}c^{\left(n\right)}(s)$, which put into Eq.~(\ref{lorentz_exact_s}) produces the perturbation corrections,
\begin{eqnarray*}
	\alpha^0:\qquad && s c^{\left(0\right)}(s)=1,\nonumber\\
	\alpha^2:\qquad && s c^{\left(1\right)}(s)=-\frac{1}{2s}c^{\left(0\right)}(s),\nonumber\\
	\alpha^4:\qquad && s c^{\left(2\right)}(s)=-\frac{1}{2s}c^{\left(1\right)}(s)+\frac{1}{2s^2}c^{\left(0\right)}(s),\nonumber\\
	\vdots\qquad \quad&&
\end{eqnarray*}
After some calculations, for LT2, $\rho\simeq|1-\alpha^2\tilde{t}^2/4|^2$; for LT4, $\rho\simeq|1-\alpha^2\tilde{t}^2/4+\alpha^4(8\tilde{t}^3+\tilde{t}^4)/96|^2$. None of them converges when $\tilde{t}\to\infty$.
\\

Likewise, if $c(s)$ is expanded around $\alpha^{-1}$, then putting $c(s)=\sum\nolimits_{n}\alpha^{-2n}c^{\left(n\right)}(s)$ into Eq.~(\ref{lorentz_exact_s}) yields
\begin{eqnarray*}
	\alpha^{-0}:\qquad && s c^{\left(0\right)}(s)-1=-\frac{1}{2}c^{\left(0\right)}(s),\nonumber\\
	\alpha^{-2}:\qquad && s c^{\left(1\right)}(s)=-\frac{1}{2}c^{\left(1\right)}(s)+\frac{s}{2}c^{\left(0\right)}(s),\nonumber\\
	\vdots\qquad \quad&&
\end{eqnarray*}
For LT(--0), $\rho\simeq|\exp(-\tilde{t}/2)|^2$; for LT(--2), $\rho\simeq|\exp(-\tilde{t}/2)+\alpha^{-2}(\tilde{t}-2)\exp(-\tilde{t}/2)/4|^2$. There is only decay but no oscillation.

\begin{widetext}
\section{Multiple-scale perturbation corrections for $N=1$ systems}
\label{appendix_n1}
One should bear in mind that $(d/d\tilde{t})^{n+1}\int{d{\tilde{t}'}(\tilde{t}-\tilde{t}')^{n}c^{(m)}(\tilde{t}')}=n!c^{(m)}(\tilde{t})$. When $N=1$, from Eq.~(\ref{integro-differential_ctqw_expand}) the iterative procedure yields
\begin{eqnarray*}
\alpha^2:\qquad &&A_{0}^{2}\frac{{{\partial }^{2}}}{\partial {{\tilde{T}}^{2}}}c^{\left( 0 \right)}=-G_0 c^{\left( 0 \right)},\nonumber\\
\alpha^4:\qquad && A_0^3 \frac{{{\partial }^{3}}}{\partial {{\tilde{T}}^{3}}}c^{\left( 1 \right)}+2 A_0^2 B_0 \frac{{{\partial }^{3}}}{\partial {{\tilde{T}}^{2}}\partial {{\tilde{\tau}}}}c^{\left( 0 \right)}=- A_0 G_0 \frac{{{\partial }}}{\partial {{\tilde{T}}}}c^{\left( 1 \right)}- G_1 c^{\left( 0 \right)},\nonumber\\
\alpha^6:\qquad &&A_0^4 \frac{{{\partial }^{4}}}{\partial {{\tilde{T}}^{4}}}c^{\left( 2 \right)}
+3A_0^3 B_0 \frac{{\partial }^4}{\partial {{\tilde{T}}^3}\partial {{\tilde{\tau}}}}c^{\left( 1 \right)}
+3A_0^2 B_0^2 \frac{{\partial }^4}{\partial {{\tilde{T}}^2}\partial {{\tilde{\tau}}^2}}c^{\left( 0 \right)}
+2A_0^3 A_1 \frac{{\partial }^4}{\partial {{\tilde{T}}^4}}c^{\left( 0 \right)}\nonumber\\
&=&-A_0 B_0 G_0 \frac{{\partial }^2}{\partial {{\tilde{T}}}\partial {{\tilde{\tau}}}}c^{\left( 1 \right)}
-A_0^2 G_0 \frac{{\partial }^2}{\partial {{\tilde{T}}^2}}c^{\left( 2 \right)}
-A_0 G_1 \frac{\partial }{\partial {\tilde{T}}}c^{\left( 1 \right)}
-2 G_2 c^{\left( 0 \right)},\nonumber\\
\alpha^8:\qquad &&A_0^5 \frac{{{\partial }^{5}}}{\partial {{\tilde{T}}^{5}}}c^{\left( 3 \right)}
+4A_0^4 B_0 \frac{{\partial }^5}{\partial {{\tilde{T}}^4}\partial {{\tilde{\tau}}}}c^{\left( 2 \right)}
+6A_0^3 B_0^2 \frac{{\partial }^5}{\partial {{\tilde{T}}^3}\partial {{\tilde{\tau}}^2}}c^{\left( 1 \right)}
+4A_0^2 B_0^3 \frac{{\partial }^5}{\partial {{\tilde{T}}^2}\partial {{\tilde{\tau}}^3}}c^{\left( 0 \right)} \nonumber\\
&&+3A_0^4 A_1 \frac{{\partial }^5}{\partial {{\tilde{T}}^5}}c^{\left( 1 \right)}
+2A_0^4 B_1 \frac{{\partial }^5}{\partial {{\tilde{T}}^4}\partial {{\tilde{\tau}}}}c^{\left( 0 \right)}
+8A_0^3 A_1 \frac{{\partial }^5}{\partial {{\tilde{T}}^4}\partial {{\tilde{\tau}}}}c^{\left( 0 \right)} \nonumber\\
&=&-A_0 B_0^2 G_0 \frac{{\partial }^3}{\partial {{\tilde{T}}}\partial {{\tilde{\tau}}^2}}c^{\left( 1 \right)}
-A_0^2 A_1 G_0 \frac{{\partial }^3}{\partial {{\tilde{T}}^3}}c^{\left( 1 \right)}
-2 A_0^2 B_0 G_0 \frac{{\partial }^3}{\partial {{\tilde{T}}^2}\partial {{\tilde{\tau}}}}c^{\left( 2 \right)}
-A_0^3 G_0 \frac{{\partial }^3}{\partial {{\tilde{T}}^3}}c^{\left( 3 \right)} \nonumber\\
&&-A_0 B_0 G_1 \frac{{\partial }^2}{\partial {{\tilde{T}}}\partial {{\tilde{\tau}}}}c^{\left( 1 \right)}
-A_0^2 G_1 \frac{{\partial }^2}{\partial {{\tilde{T}}^2}}c^{\left( 2 \right)}
-2 A_0 G_2 \frac{\partial }{\partial {\tilde{T}}}c^{\left( 1 \right)}
-6 G_3 c^{\left( 0 \right)}.\nonumber\\
\vdots\qquad \quad&&
\end{eqnarray*}
The $\alpha^6$ and $\alpha^8$ order correction terms are of more interest because they further set constraints on $A_1$ and $B_1$, by
\begin{equation*}
A_1=\left(\frac{3}{8}\frac{G_1^2}{G_0^3}-\frac{G_2}{G_0^2}\right) A_0
\qquad\text{and}\qquad
B_1=\left(-\frac{G_1^2}{G_0^3}+4\frac{G_2}{G_0^2}-6\frac{G_3}{G_1G_0}\right) B_0.
\end{equation*}
\end{widetext}

\bibliography{MultiScale}

\end{document}